\documentclass[preprint,12pt]{elsarticle}

\usepackage{amssymb}
\usepackage{amsmath}

\usepackage{algpseudocode}
\usepackage{algorithm}
\usepackage{mathtools}

\usepackage{subfig}
\usepackage{graphicx}

\usepackage{hyperref}

\usepackage{xcolor}

\journal{Internet of Things}

\begin{document}

\begin{frontmatter}



\title{BDPC: Controlling Application Delay in 6TiSCH networks for the Industrial Internet of Things}


\author[inst1]{Lucas Aimaretto}

\affiliation[inst1]{
            organization={Facultad de Ciencias Exactas, Físicas y Naturales, Universidad Nacional de Córdoba},
            state={Córdoba},
            country={Argentina},
            }

\author[inst2]{Diego Dujovne}

\affiliation[inst2]{organization={Escuela de Informática y Telecomunicaciones, Universidad Diego Portales},
            state={Santiago},
            country={Chile}}

\begin{abstract}

One of the essential requirements of wireless industrial Internet of Things (IoT) systems is to have an extremely high packet delivery rate, generally over 99.9\% and comply wih realtime deadline constraints. In industrial IoT networks, packets arriving after the deadline become part of packet loss and lose meaning when they arrive late. However, currently available industial IoT proposals aim to minimize End-to-End delay without taking into account simultaneous realtime and reliability constraints. In this paper, we propose a new mechanism, called BDPC (Bounded Delay Packet Control) to tackle this challenge. BDPC combines the knowledge of a node's traffic delay to the destination (root) with the time budget of a data packet traversing the industrial IoT network, to allocate network resources to comply the system maximum delay requirements using an adaptive and distributed algorithm. Unlike the general aim to minimze end-to-end delay, we propose that data packets must arrive before the deadline, but not faster. Our results show, for example, that by using BDPC, the number of packets arriving before the deadline can be improved more than 2.6 times compared to the case when using the default Minimal Scheduling Function from the standard. As a further advantage, BDPC involves minor modifications to the 6TiSCH protocol stack, which makes it compatible with current implementations.

\end{abstract}


\begin{highlights}
\item Scheduling Function to address resource allocation considering the delay to root node and the consumed time budget by data packets. This allows for the network to cope with the application's $deadline$ requirements.
\item The resource allocation is addressed on the opposite direction of data traffic flow.
\end{highlights}


\end{frontmatter}


\section{Introduction}
\label{section:intro}

 There is an increasing number of industrial production processes around the world which are transitioning to the era of Industry 4.0. This paradigm change comprises a complete redesign to achieve the convergence of Operational Technologies (OT) for control processes with Information Technologies (IT) from data transmission and processing instances. The Industrial Internet of Things (Industrial IoT) is the response to this evolution. One of the Industrial IoT key enablers is the construction of an integrated network  to support heterogeneous traffic patterns, composed by different data flows covering a range from best effort traffic to time-sensitive realtime applications traffic.\\

\begin{figure}
	\centering
	\includegraphics[scale=0.6]{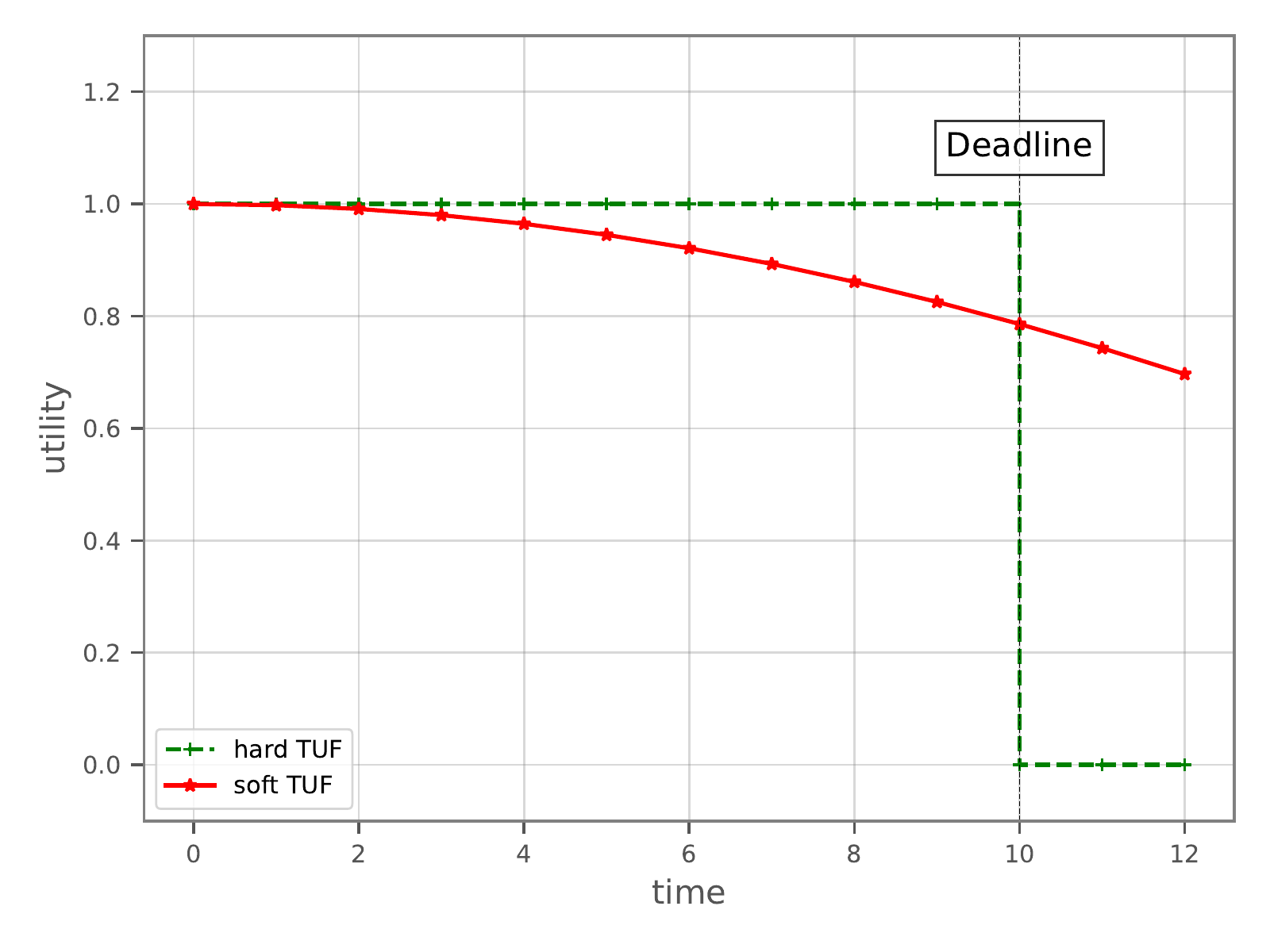}
	\caption{Hard and Soft Time Utility Functions. For the hard case, if the task is completed after the maximum deadline, the outcome is discarded.}
	\label{fig:tuf}
\end{figure}

 In realtime systems, time-constrained tasks are defined as a function of its maximum limit, called the \emph{deadline}. Time Utility Functions (TUFs) define two kinds of deadlines \cite{ryu2005urgency}: hard deadlines and soft deadlines. The hard deadline case establishes a constraint on the maximum limit in which the task must be completed, while if the condition is a soft deadline, the task is less strict on the arrival time and can be completed within a wider period. The hard deadline case is associated with real time applications, such as industrial control systems. Thus, for the hard deadline case, if any task is completed after the maximum deadline, the outcome is discarded, even if the outcome is correct. Hard and soft TUFs behavior can be observed in Fig. \ref{fig:tuf}.

In addition to the hard deadlines required by real-time applications, not only end-to-end delay must be compliant, but also variability around the target packet arrival time (jitter) must be reduced.  Packet arrival time variability introduces noise and unexpected dead times in control systems and hence it must be bounded to reduce the former effects \cite{sunori2017dead, ogunnaike1994process}.

Realtime systems traffic have several strict constraints. Among them, first, it is fundamental to guarantee an extremely high packet delivery rate \cite{dujovne20146tisch}, and second, to satisfy realtime performance, data packets must also arrive to destination within a limited time frame.  As a matter of fact, a packet which is delayed after the expected  application  deadline  must be  discarded by the application:  even if the packet arrived correctly at destination, it is  considered lost since the payload is not valid anymore. \\

 Industrial IoT networks are based on both wired and wireless communications infrastructure. Wireless infrastructure provides greater flexibility, maintenance and scalability than wired infrastructure, for fixed location measurement and actuation nodes. However, realtime restrictions compliance using wireless infrastructure is still an open challenge.
Currently, the standardized Architecture of Deterministic Networks \cite{RFC8655}  describes both  minimum Packet Delivery Rate (PDR) and time-bounded data packet delivery, or \emph{bounded latency}  as requirements  for wired solutions, while wireless Industrial IoT has been following  an adjacent  path, specifically within the scope of the Reliable and Available Wireless standardization group (RAW) from the Internet Engineering Task Force (IETF). 

Moreover, while realtime restrictions are fundamental requirements to enable robust and reliable Industrial IoT solutions, Industrial IoT networks must make rational resource utilization in order to keep network lifetime and at the same time, make proportional use of transmission opportunities. In this work, we propose a new distributed resource allocation algorithm to dynamically allocate transmission resources along the Industrial IoT networks for realtime-constrained flows. Unlike other proposals which try to minimize delay to reach the destination (root) node, our work aims to equalize packet arrival times without increasing resource usage.  

\begin{figure}
	\centering
	\includegraphics[scale=0.7]{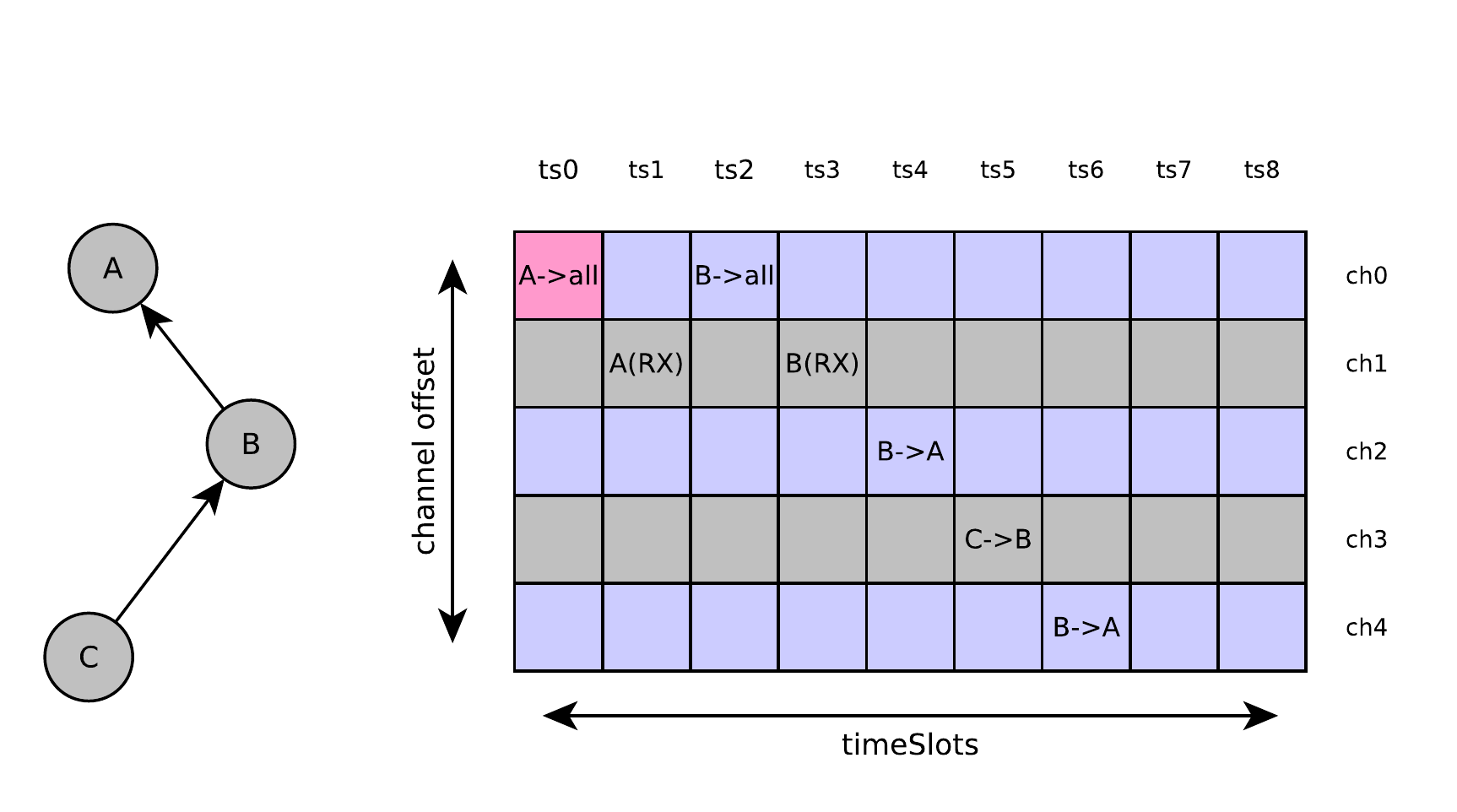}
	\caption{In the example, Node B will transmit to node A, when timeSlot 4 occurs, doing so on channelOffset 2. The Cell on coordinate [0,0] is called the Minimal-Cell, which is used for broadcast traffic.}
	\label{fig_intro_net_schedule}
\end{figure}

One of the most advanced wireless Industrial IoT implementations is the Time Slotted Channel Hopping (TSCH) mode of the IEEE 802.15.4 standard, which defines how nodes reserve resources according to a channel-timeslot schedule matrix, called slotframe. The slotframe matrix determines a number of cells, where a pair of nodes over a communication link can exchange packets. TSCH is based on a multihop topology, specifically designed to offer reliability by incorporating path redundancy. 
 Fig. \ref{fig_intro_net_schedule} shows an example configuration of the TSCH matrix for a 3-node network. The schedule establishes a specific task for each cell: transmit, receive or sleep. Each cell represents a transmission opportunity between nodes.

Additionally, TSCH nodes first start a neighbor discovery process to obtain a list of reachable neighbors, which further relay the data packets towards a root node. The destination (root) node is a fixed node in the network which operates as a border router which has access to other networks or even the global Internet.

\begin{figure}
	\centering
	\includegraphics[scale=0.7]{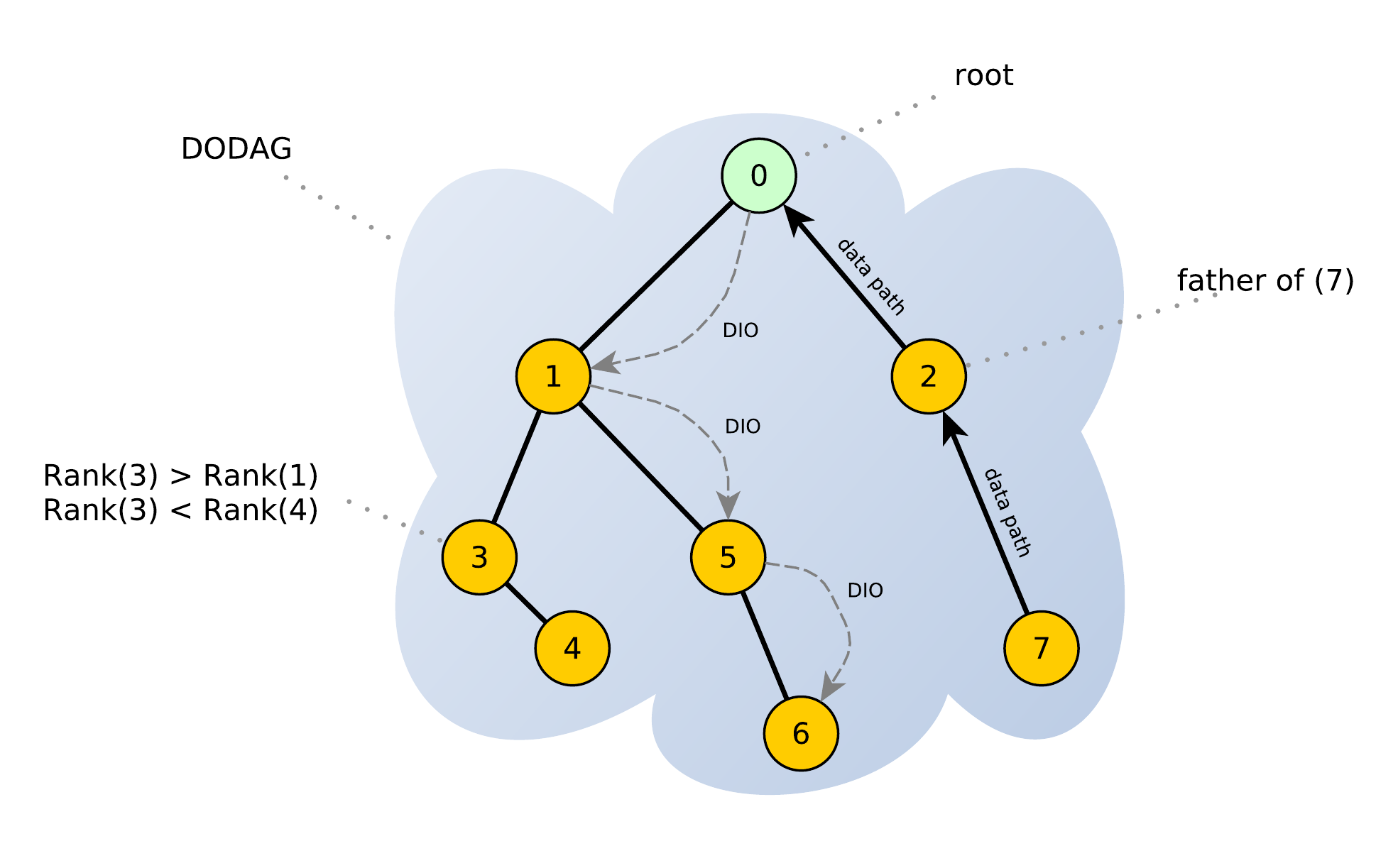}
	\caption{All the nodes forward traffic using a multihop topology to the destination (root) node.  In the example Rank(3)$>$Rank(1) and hence, node 1 becomes the parent of node 3. The Rank value is shared among nodes using the DIO messages.}
	\label{fig_intro_dodag}
\end{figure}

The IPv6 over the TSCH mode of IEEE 802.15.4e (6TiSCH) stack relies on a routing protocol that allows nodes to build a logical infrastructure to let the traffic reach the destination (root) node. This protocol is called IPv6 Routing Protocol for Low-Power and Lossy Networks (RPL) \cite{rfc6550}. RPL uses messages that lets nodes discover neighbors and build the routing structure. The routing algorithm is called Destination Oriented Directed Acyclic Graph (DODAG) and it is based on an individual Rank value, which is shared by specific DIO messages (DODAG Information Objects). The logic behind the algorithm relies on the fact that - except the destination (root) node - each node will always become child of a parent node, based on the Rank value comparison between the nodes, as it can be appreciated in Fig. \ref{fig_intro_dodag}. A child node will select another node as a parent only if the Rank value of the parent candidate node is lower than the child node Rank.

 Finally,  the 6TiSCH stack establishes the mechanisms for resource allocation, leaving to the Scheduling Functions (SFs) the implementation of the algorithms that decide when and how to allocate resources \cite{rfc8180}.

Scheduling Functions (SF) are the core of the allocation intelligence in 6TiSCH-based IoT networks.  They can be centralized or distributed, each of those with their different advantages and disadvantages. When centralized, there is a central entity called the Path Computation Element (PCE) which calculates the optimal schedule to be deployed on the TSCH matrix. The PCE has a complete view of the network and hence the schedule is optimally optimized to cope with the network requirements. However, when the network is a Low Power and Lossy Network (LLN), the PCE reactivity to variable link conditions can be far from optimal \cite{urke2021survey}.

\begin{figure}
	\centering
	\includegraphics[scale=0.6]{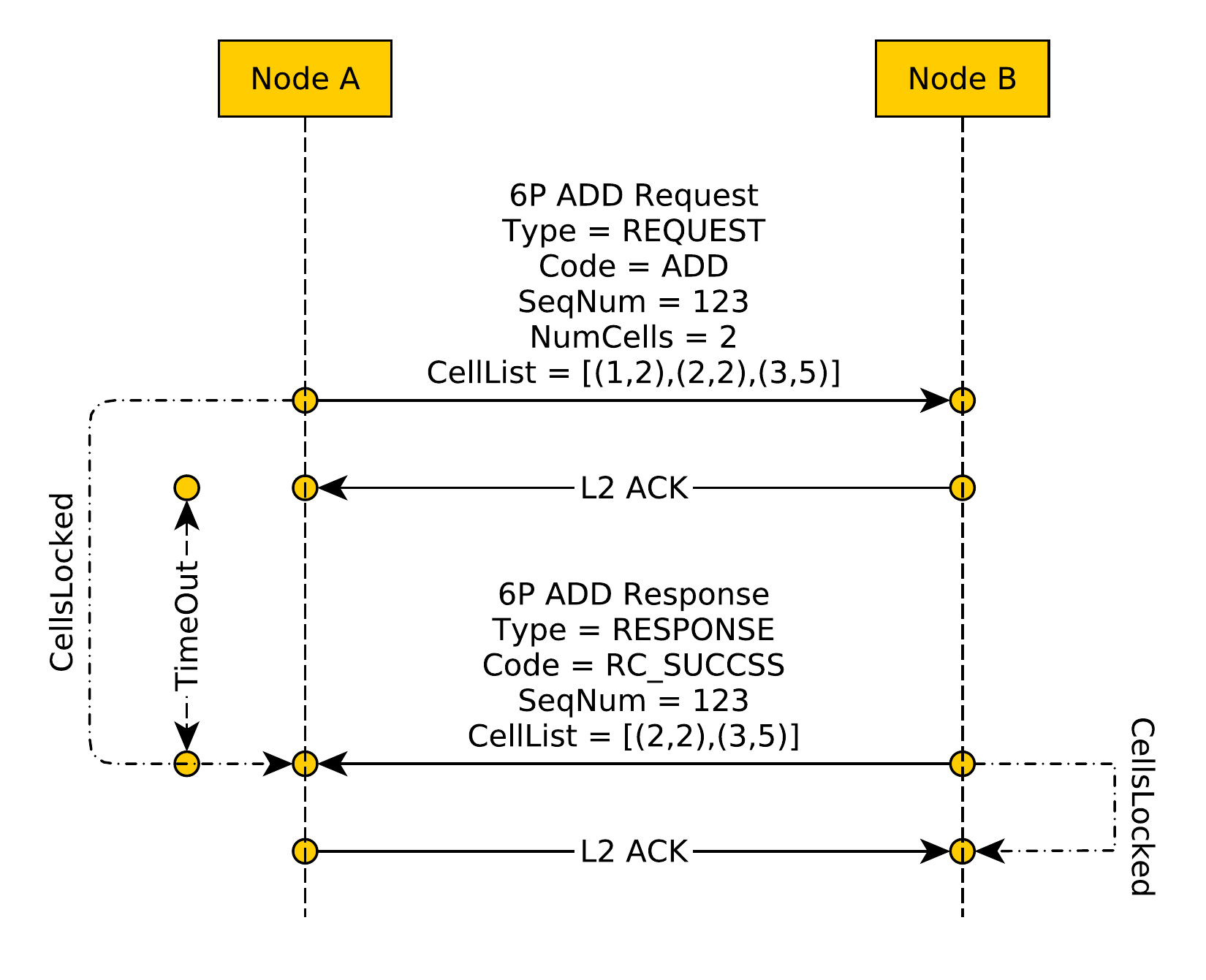}
	\caption{6P two-way handshake for cell assignment.}
	\label{fig:6p_2way}
\end{figure}

On the other hand, when the SF is distributed, a pair of nodes from the network exchange messages to agree on the resource allocation using the 6P protocol \cite{rfc8480}. By using a set of messages, the pair of nodes can add or remove cells from the schedule. Fig. \ref{fig:6p_2way} shows the schematic representation of 6P when exchanging messages between a pair of nodes, to add cells to the schedule.

SFs implement a resource allocation algorithm designed to improve the network behavior by targeting specific network parameters such as delay, power consumption, network lifetime and packet delivery rate. Among the wide range of proposed SFs \cite{urke2021survey}, only a fraction of them aim to reduce the end-to-end delay to the minimum possible.  However, the latter SFs minimize end-to-end delay with no target deadline nor concurrent PDR target values complying with realtime constraints. To the best of our knowledge, there is currently no available SF designed to maximize the Packet Delivery Rate (PDR) and simultaneously fulfill a certain application-defined $deadline$.

In order to provide a solution to wireless industrial IoT realtime constrained systems, we propose the BDPC Scheduling Function (Bounded Delay Packet Control). BDPC deals with realtime systemss with specific hard deadline simultaneously with packet delivery rate compliance requirements. BDPC solves this problem by combining upstream and downstream network information to allocate network resources towards the source, i.e. on the child node link, thus adapting the number of transmission opportunities in a hop-by-hop basis. As a consequence, BDPC enables the highest number of packets possible to reach the destination before the expected deadline. As a matter of fact, the fundamental difference between BDPC and current delay minimization proposals such as \cite{rfc9033,ldsf, resf, llsf} is the main goal: BDPC aims to simultaneously guarantee deadline compliance for real-time traffic with an adjustable Packet Delivery Rate without spending resources on extreme delay minimization. Furthermore, BDPC is agnostic either to traffic patterns or topology structure and respects data flow isolation. 

\bigskip

In order to solve the need for deadline and maximum packet loss compliance in wireless industrial IoT systems, our work provides the following contributions:

\begin{itemize}
    
    \item Given network realtime constraints in wireless industrial IoT control systems, BDPC dynamically configures the network resources to comply with the predefined packet arrival deadline simultaneously with a bounded packet delivery rate. 
    \item Given that Jitter may generate instability in wireless industrial IoT control systems, BDPC bounds packet delay arrival variability.
    \item Given that wireless industrial IoT control systems are prone to be disrupted by energy depletion, BDPC provides an adjustable trade off between power consumption and realtime constraint compliance.

\end{itemize}

The rest of the paper is organized as follows: Section \ref{section:motivation} provides the state of the art for this work; section \ref{section:propuesta} presents the BDPC algorithm and the full compatibility with current standards. In section \ref{section:results}, we present the results based on simulation work; and finally in section \ref{section:conclusion} we conclude our work.
\section{State of the art}
\label{section:motivation}


To the best of our knowledge, there are no currently published SFs for wireless industrial IoT control systems which are aimed to comply with realtime constraints, comprising the packet arrival deadline and bounded packet delivery rate. The closest approach we have identified are those centered on those SFs which aim to minimize packet delays, where we can find the following proposals. 

First, MSF \cite{rfc9033} is the Minimal Scheduling Function, which is the full-featured standard SF used by default within the 6TiSCH stack. MSF recommends increasing the number of cells in the data plane to decrease latency, at the cost of power consumption increase.


Second, the Low Latency Scheduling Function (LLSF) \cite{llsf} was designed to decrease end-to-end latency in 6TiSCH multi-hop networks, using a \emph{daisy-chain} approach.
The transmission sequence is defined by a succession of allocated timeSlots incrementally shifted in time for each node in the network. Once a packet is received in a RX timeSlot, it is automatically transmitted to the next node in the following TX timeSlot.
The results in \cite{llsf} show that the end-to-end delay is 82.8\% lower than using SF0 \cite{draft-ietf-6tisch-6top-sf0} for a 5-hop linear topology.



Third, the Low-latency Distributed Scheduling Function (LDSF) \cite{ldsf} algorithm is aimed to minimize the end-to-end delay while providing high reliability. LDSF is able to generate a schedule for a large number of transmissions, also managing the buffering delay when oversubscribed cells are chained. In order to achieve this goal, LDSF divides a long slotframe into small blocks that repeat over time. Each node then selects the right block corresponding to its hop distance to the border router which minimizes delay.


The paper introduces concept of primary and ghost cells. The primary cells correspond to the earliest expected reception time of a data packet from the previous hop. To guarantee network reliability, LDSF allocates cells to enable  a retransmission opportunity in the event of a transmission failure of a data packet during the primary cell. The ghost cells are allocated at the same timeSlot offset in subsequent subFrames. 

By performing simulations, LDSF delivers a $PDR_{e2e}$ above 98.5\% with a network lifetime of 4 years, for a traffic flow of 1 packet every 30s. In this scenario, the end-to-end delay is 250ms.


Finally, the Recurrent Low-Latency Scheduling Function (ReSF) \cite{resf} is a SF that focuses on optimizing application delay with a periodic traffic pattern.

This function uses a distributed method to allocate resources through 6P \cite{rfc8480}, focusing on building a resource reservation mechanism that minimizes recurring communication latencies, aimed only towards IoT systems where devices perform periodic data transmission. ReSF assumes that the source node is aware of the traffic period pattern. With this information, ReSF is able to generate a recurring path, which consists of cells that are activated only when traffic is expected and deactivated otherwise. ReSF allows nodes to make reservations that are based on the Absolute Slot Number (ASN) where it started generating data, the ASN where it stops generating data and the repetition period. This information is sent to the next hop until it reaches the destination, thus reserving resources for each node along the path.  

By performing simulations, ReSF shows that the latency for a network size of 25 elements and a traffic flow of 1 packet per minute is close to 0.5 seconds with a $PDR_{e2e}$ of 100\%.

Although the former proposals describe SFs aimed towards or minimizing delay, their goal is not to comply with wireless industrial IoT control systems realtime constraints. First, the former proposals do not aim to fulfill a maximum delay goal corresponding to the $deadline$: This is expected in a realtime environment, for example, if a wireless industrial IoT application  running on a 6TiSCH-based stack specifies a deadline for packet arrival times.  Second, the former proposals do not consider the effect of jitter of the arriving packets in industrial IoT control systems, which increments packet loss for packets arriving after the $deadline$ has expired. And third, even though the former proposals provide a high end-to-end packet delivery rate ($PDR_{e2e}$), they do not take into account that value for wireless industrial IoT control application $deadlines$, considering that the packets arriving after the deadline will be discarded \cite{draft-ietf-raw-architecture}.

\section{Proposed Solution}
\label{section:propuesta}

Resource scheduling algorithms for the Industrial Internet of Things control systems must satisfy the realtime requirements for each data flow traversing the network nodes. Scheduling Functions aimed towards minimizing delay generally achieve this goal as a global process, where no intermediate packet delay information is taken into consideration to make a decision in a node to node basis. We propose to use the packet time budget, plus the estimation of a node's delay to the destination, to allow BDPC to allocate resources,  in the destination to source direction, and adapt to the dynamic data flow changes.

\subsection{Time budget estimation}

\begin{figure}[!hb]
	\centering
	\includegraphics[trim={0 10cm 0 2cm}, scale=0.6]{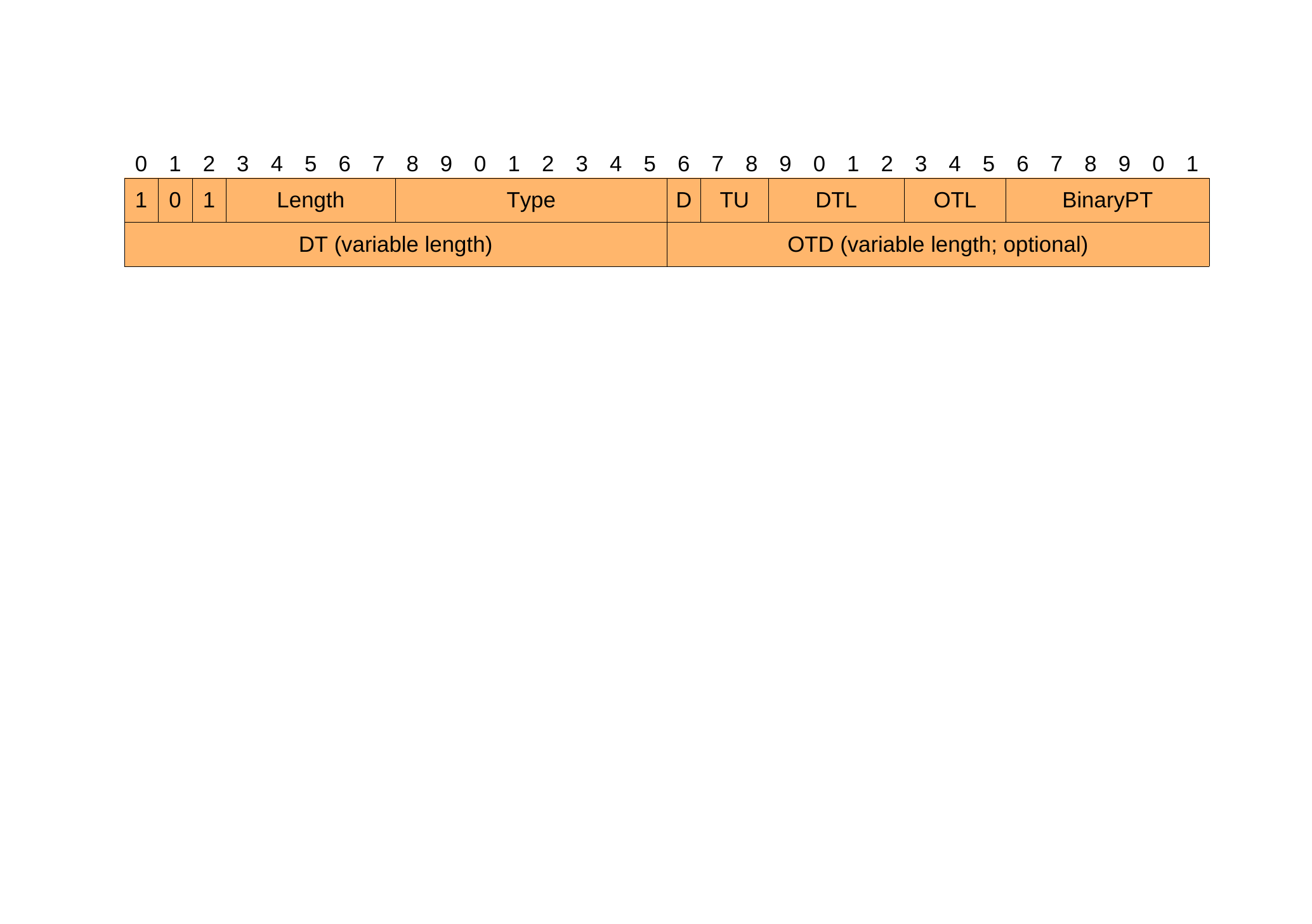}
	\caption{6LoRHE - Routing Header with timing information. The timing information is crucial to calculate the time budget of a data packet. The $deadline$ is carried inside the DT field. }
	\label{fig:6lorhe}
\end{figure}

 The $deadline$ value can be included in a standard data packet:  The Packet Delivery Deadline Time field in the Routing Header for IPv6 over Low-Power Wireless Personal Area Networks \cite{rfc9034} is the current standard field where the deadline time for the data packets can be propagated from origin to destination. This header (Fig. \ref{fig:6lorhe}) is inserted at the Network layer (as defined on the 6LoWPAN standard  \cite{thubert2016ipv6}) and includes, among other fields, the packet origin time and the maximum delivery time of the packet.

For example, for a 6TiSCH network with 10ms timeSlots, if a source node has a number of enqueued data packets ready to be transmitted, the current Absolut Slot Number (ASN) is 54400, which represents the $originTime_{packet}$. The maximum tolerable sytem delay, $maxDelay$, is 1s. The $deadline$ of the application, is calculated as follows:

\begin{equation}
	\begin{split}
		deadline & = originTime_{packet} + \texttt{maxDelay}/10ms \\
		deadline & = 54400 + 100 \\
		deadline & = 54500
	\end{split}
	\label{eq_deadline_calculation}
\end{equation}

\bigskip

Consequently, the application data packet $deadline$ value to be included in the \texttt{DT} field is $54500$ (Fig. \ref{fig:6lorhe}), and the maximum supported delay is 100 timeSlots.

If an intermediate node receives a packet and analyzes the 6LoRHE header, that node can estimate if the packet is still within the delivery $deadline$, by calculating the $timeLeft$, which is the difference between the $deadline$ carried by the packet, and the current time. To follow the former example, if the  intermediate node receives the packet at ASN 54450:

\begin{equation*}
	\begin{split}
		timeLeft_{asn} & = deadline_{packet} - currentTime_{asn} \\
		timeLeft_{asn} & = 54500 - 54450 \\
		timeLeft_{asn} & = 50 \\
		timeLeft_{seg} & = 0.5
	\end{split}
\end{equation*}

\bigskip

The delay information carried by the packet is not enough for BDPC to trigger the cell allocation process. A negative $timeLeft$ value means that the time budget has been consumed,  so the packet will not comply with the $deadline$, while if the $timeLeft$ is positive, the received packet has 0.5s remaining to reach its destination. However, at this point in the multihop network, it is still unknown whether this positive $timeLeft$ is enough to reach the root node. 

Thus, if an intermediate node  can become  aware of its own estimated delay to the destination node, as well as the $timeLeft$ value for the data packet, BDPC will be able to make a cell scheduling decision for the traversing packet flow.  This last issue is solved by means of a new variable called $d2r$.


In this proposal, a node that generates DIO messages includes  additional information to allow the child nodes to be aware of the delay to the destination (root) node. When a child node receives a DIO message with timing information from its parent node, it updates the $d2r$ variable (delay-to-root) in the internal $DB_{dio}$ database. Timing information in DIOs must only be considered by a node when originated from a parent node to keep consistent values.


\subsection{System Model}

We define the single path end-to-end data packet delay as the difference between the arrival time of a data packet at the destination (root) node, and the departure time from a source node, either a terminal or an intermediate node: $d=t_{rx} - t_{tx}$. 







A source node that generates a packet towards the destination (root) node, uses Eq. (\ref{eq_deadline_calculation}), inserts the $deadline$ field in the 6LoRHE (Fig. \ref{fig:6lorhe}) and delivers the data packet to the next-hop in the next available timeSlot. The closer the next available timeSlot, the shorter the incremental delay will be.

Additionally, both the destination (root) node and every intermediate node, calculate the $latePaqs$ variable as well, as defined by Eq. (\ref{eq_latePaqs}). Incoming packets at the intermediate node, are identified by the source MAC address. Hence, there is a different $latePaqs$ value per each source MAC for every children of an intermediate node. The $latePaqs_{MAC}$ variable is calculated using Alg. \ref{alg:delayed}.

\begin{equation}
latePaqs_{MAC} = f(deadline,d2r) =  \frac{delayed}{delayed+inTime}
\label{eq_latePaqs}
\end{equation}

\bigskip


In a 6TiSCH network, the TSCH slotFrame is composed of cells that can be of type \texttt{SHARED}, \texttt{TX}, \texttt{RX}, or a combination of them. According to \cite{rfc9033}, the delay can be reduced by adding \texttt{RX} cells from a parent node to a child node into the schedule so data packets will have more transmission opportunities while traversing the path towards the destination (root) node.  BDPC aims to increase the $inTime$ packet count and reduce the $delayed$ packet count by adjusting hop-to-hop allocated resources dynamically in the destination to source direction.



BDPC can be parametrized by defining the upper threshold that triggers  cell allocation, as defined in Alg. \ref{alg:slots}. This parameter is called \texttt{sfMax}. On the other end, BDPC also defines \texttt{sfMin} as the lower threshold. If $latePaqs$ is below \texttt{sfMin}, cell removal will be triggered to save energy. This behavior is part of Alg. \ref{alg:slots}.

\subsection{BDPC Scheduling Function}

\begin{figure}
	\centering
	\includegraphics[scale=0.7]{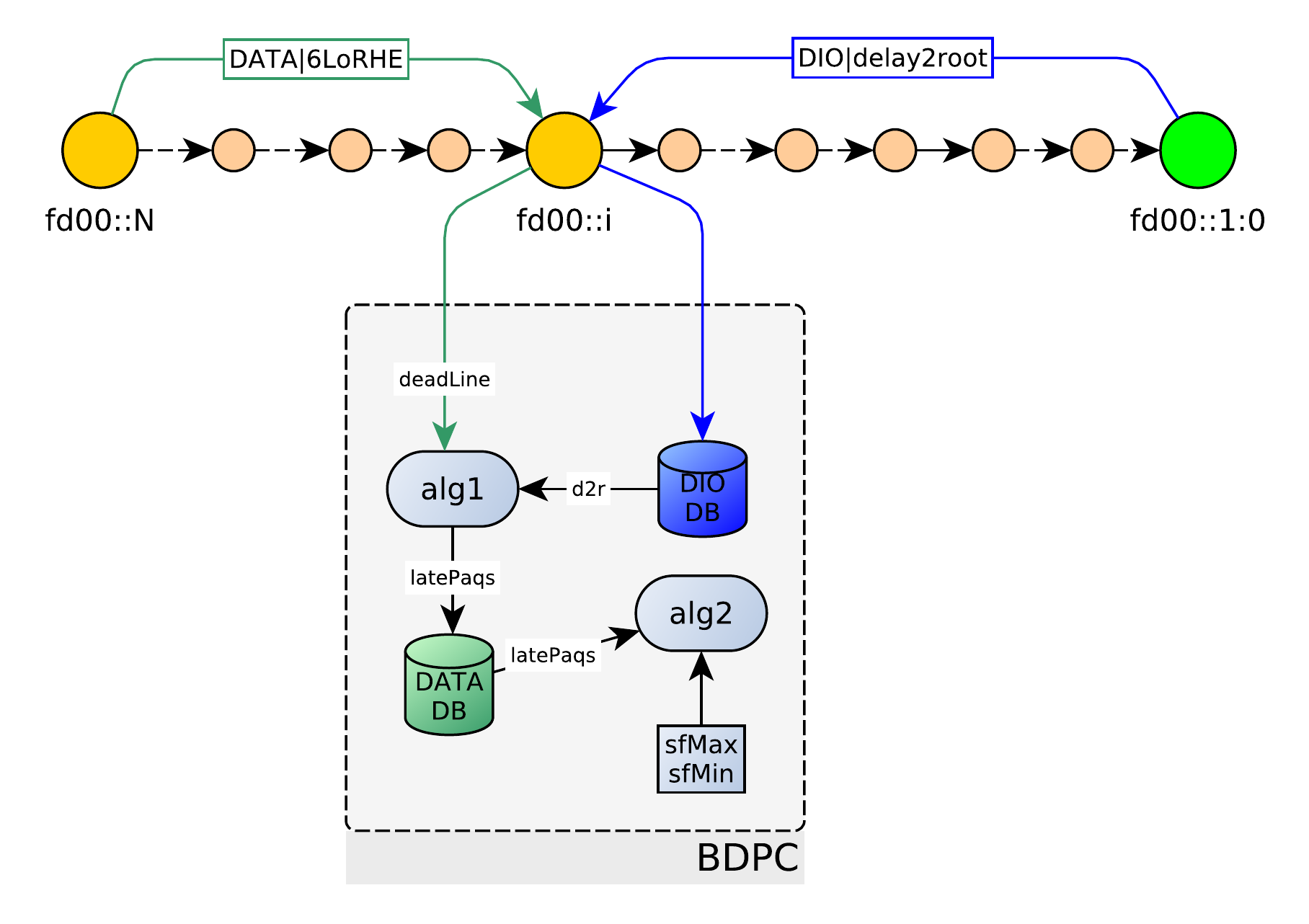}
	\caption{BDPC system at an intermediate node. Each node measures its own delay to the root node.}
	\label{fig:DIO_DATA_Topology}
\end{figure}

\begin{figure}
	\centering
	\includegraphics[scale=0.8]{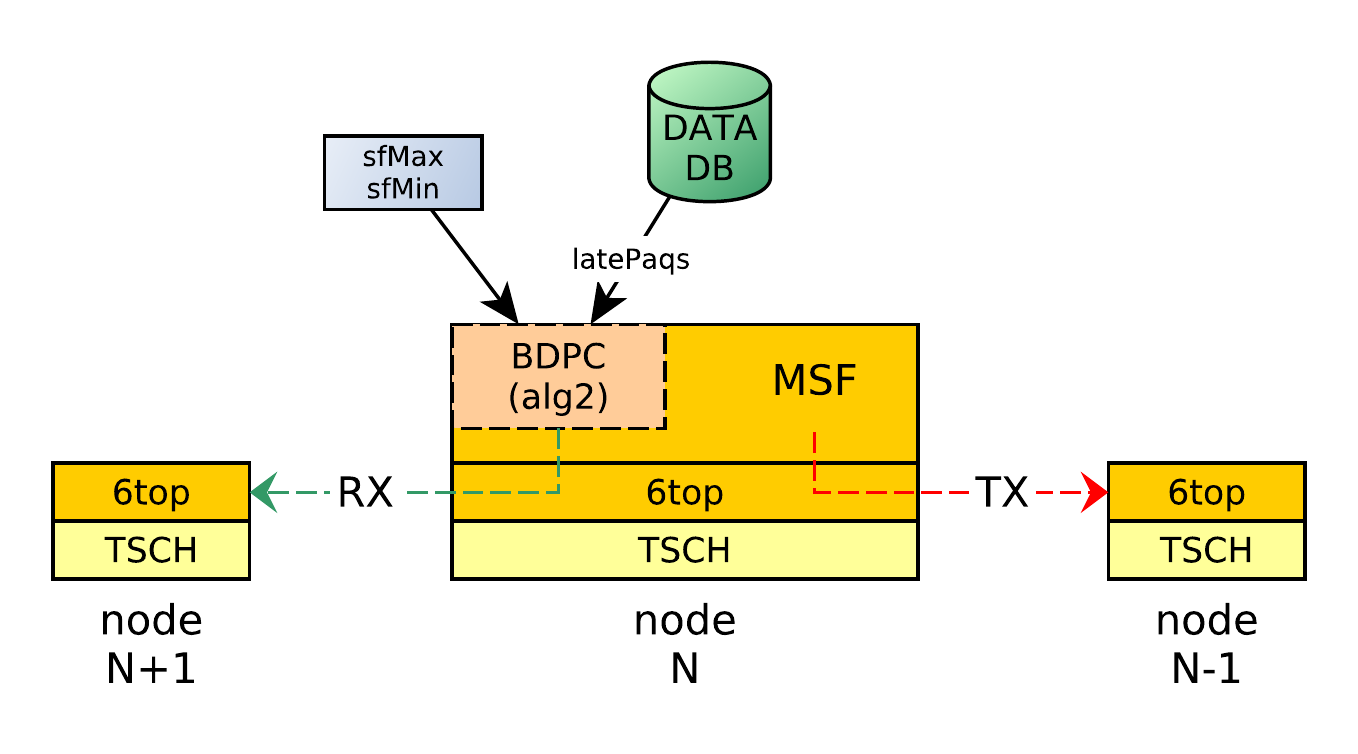}
	\caption{BDPC and MSF, at an intermediate node. BDPC will add RX cells to the previous hop (Pre-Hop), while MSF will add TX cells to the Next-Hop.}
	\label{fig_MSF_BDPC_stack}
\end{figure}

BDPC is a system based on a simple and modular architecture design contained within each of the network intermediate nodes as a Scheduling Function (Fig. \ref{fig:DIO_DATA_Topology}). It is designed to run as part of the 6TiSCH protocol stack and interacts with 6P and the TSCH MAC layer (Fig. \ref{fig_MSF_BDPC_stack}). BDPC uses two complementary algorithms: Algorithm \ref{alg:delayed} calculates the rate of packets arriving late, called $latePaqs$ and Algorithm \ref{alg:slots} uses this value as an input to trigger the cell allocation process.  
Historical values of the variables are stored in databases.
$DB_{data}$ and $DB_{dio}$ databases store data packet statistics. Finally, delay to destination node (in this case the root node) measurements are transported by the DIO RPL \cite{rfc6550} packet. 

Algorithm \ref{alg:delayed} estimates the rate of packets arriving late, $latePaqs$. When an intermediate node's SF receives a data packet, Algorithm \ref{alg:delayed} calculates the $timeLeft$ of the packet and compares it  with $d2r$, which is the delay from the intermediate node to the destination (root) node. The result from the comparison is used to evaluate if the remaining time value, $timeLeft$, is enough to reach the destination (root) node.  If the remaining time is not enough to reach the destination (root) node, the $delayed$ variable is increased to record a new packet missing the deadline; otherwise, $inTime$ is increased, to record a new packet with enough remaining time to reach the deadline. $latePaqs$ is calculated as the ratio of delayed packets to the total number of packets for a timeframe. $latePaqs$ provides information to intermediate nodes about each child's current delayed packet ratio arrival statistics.

 At an intermediate node, each child is identified by the source MAC address of the receiving data frames. Hence, there is a version of $latePaqs$ for each of the children a node has.



$latePaqs$ provides an intermediate node the instantaneous ration of delayed packets to that will not arrive before the deadline to the destination (root) node, to the total number of arrived packets at the destination. A packet is considered a late packet when the estimated time lapse to arrive at the root node from an intermediate node, is larger than the packet's own remaining time. $latePaqs$ accumulates all the packets arriving fulfilling the former condition from each of its children. Consequently, $latePaqs$ informs an intermediate node the instantaneous rate of packets that are already late at that node.

As such, $latePaqs$ becomes the fundamental element of BDPC: It is calculated by algorithm \ref{alg:delayed} and it is used as an input to the Scheduling Function to define if \texttt{RX} cells must be added or removed to the appropriate child (Fig. \ref{fig_MSF_BDPC_stack}).

 On the oher hand,  Algorithm \ref{alg:slots} uses $latePaqs$, and compares it against the \texttt{sfMax} and \texttt{sfMin} thresholds, which are external parameters to configure the behavior of the SF. Using 6P, Algorithm \ref{alg:slots} performs one of these possible options: 

\begin{itemize}
    \item add a \texttt{RX} cell towards the previous node;
    \item remove a \texttt{RX} cell towards the previous node.
\end{itemize}

 So, depending on the outcome of the comparison between $latePaqs$ and the parameters \texttt{sfMax} and \texttt{sfMin}, a cell can be added or removed from the schedule, in the parent-child direction. 


\begin{algorithm}
\caption{Estimation of $latePaqs$}\label{alg:delayed}
\begin{algorithmic}
\medskip
\State $timeLeft = deadline_{packet} - currentTime_{asn}$

\State    $d2r  \xleftarrow[]{get} DB_{dio}$
\State $inTime  \xleftarrow[]{get} DB_{data}$
\State $delayed \xleftarrow[]{get} DB_{data}$

\medskip

\If{$timeLeft >= 0$ and $timeLeft >= d2r$}
    \State $inTime = inTime + 1$
\Else
    \State $delayed = delayed  + 1$
\EndIf

\medskip

\State $latePaqs = delayed / (delayed + inTime)$
\medskip
\State $latePaqs \xrightarrow[]{store} DB_{data}$
\State $inTime \xrightarrow[]{store} DB_{data}$
\State $delayed \xrightarrow[]{store} DB_{data}$

\end{algorithmic}
\end{algorithm}

\begin{algorithm}
\caption{Cell assignment to the PreHop}\label{alg:slots}
\begin{algorithmic}
\medskip
\State $latePaqs \xleftarrow[]{get} DB_{data}$

\medskip
\If{$latePaqs >= \texttt{sfMax}$}
    \State $addCell(PreHop)$
\ElsIf{$0 <= latePaqs$ and $latePaqs <= \texttt{sfMin}$}
    \State $delCell(PreHop)$
\EndIf
\end{algorithmic}
\end{algorithm}

\subsection{Cell assignment towards the source node (PreHop)}

The typical allocation process consists on the assignment of a \texttt{TX} cell from a child to its parent. However, BDPC allocates cells in the opposite direction: towards the source node. We define PreHop to name the node which is before the current node in the source-destination direction. BDPC initiates a \texttt{RX} cell request to each child, where the child is identified by the source MAC address of the transmitted data packets. Consequently, each parent node can seamlessly manage cell allocation among the children nodes independently.

\begin{figure*}
	\centering
	\includegraphics[scale=0.4]{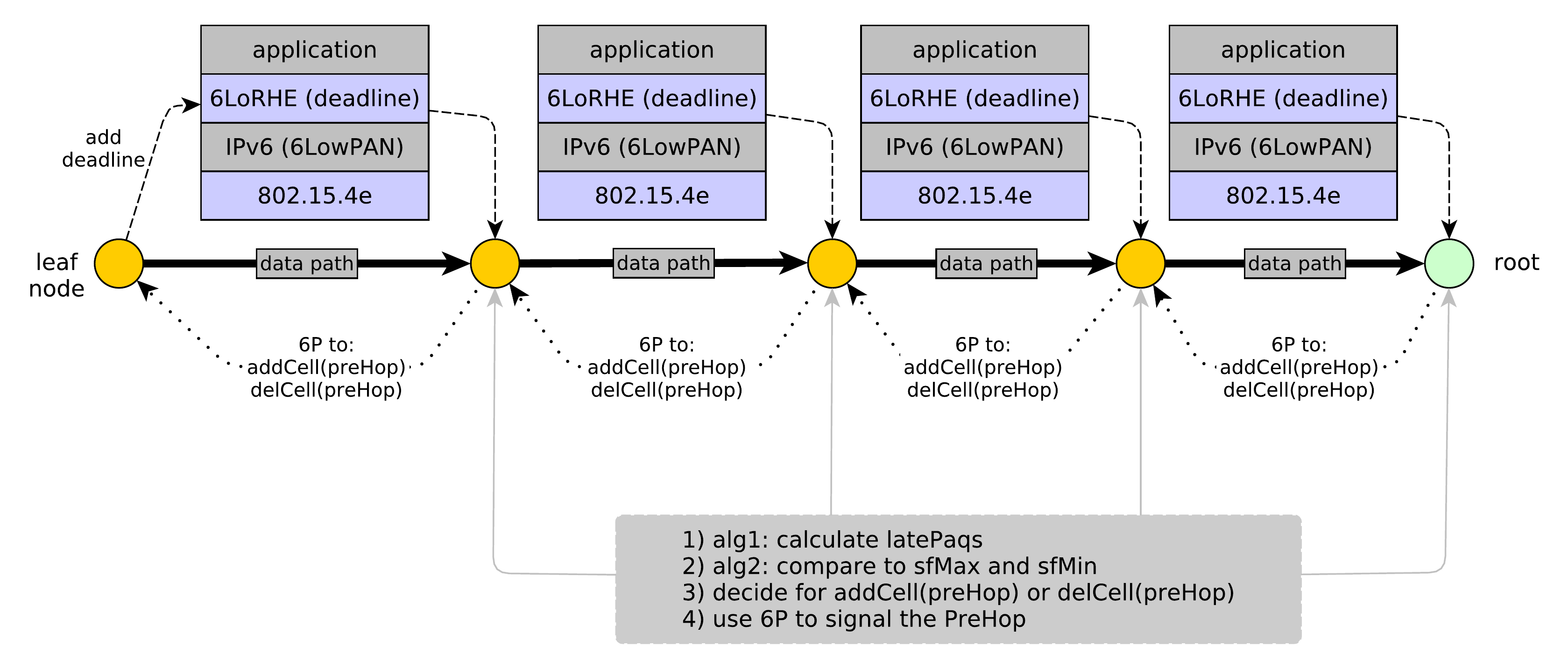}
	\caption{
	Cell allocation procedure on the parent-child direction (Pre-Hop) controlled by algorithms \ref{alg:delayed} and \ref{alg:slots}. 
	}
	\label{fig:celdas_pre_hop_topo}
\end{figure*}

Fig. \ref{fig:celdas_pre_hop_topo} shows the allocation procedure to the PreHop along the network. The transmitted IP packet includes a field inside 6LoRHE with the $deadline$ value required by the application. The packet departs from the source node towards the destination (root) node. On each intermediate node, BDPC algorithms \ref{alg:delayed} and \ref{alg:slots} process the $deadline$ value as one of the inputs. Depending on the wireless industrial IoT network conditions, algorithm \ref{alg:delayed} and algorithm \ref{alg:slots} decide either to trigger or not a new allocation process to add or remove a \texttt{RX} type cell in the PreHop direction.
The addition of a \texttt{RX} type cell to the PreHop automatically instructs the local TSCH layer of the PreHop to install a \texttt{TX} type cell towards the next-hop: The same cell is labeled RX for the parent node and TX for the child node.

The assignment process uses the 6P \cite{rfc8480} protocol, which is the same mechanism that MSF is already using when adding or removing cells towards the next-hop. Both MSF and BDPC use the same two-way 6P standard handshake for the assignment. Every time a node triggers a cell allocation process, the node starts an exchange with a 6P request to its neighbor (if MSF, the next-hop; if BDPC, the pre-hop). The client will offer a list of possible cells, out of which the server will select the ones that fit the slotFrame availability and confirm back. 

The cell allocation from the parent to the child node (\text{Pre-Hop}), managed by algorithm \ref{alg:slots} interacts seamlessly with the cell allocation mechanism to the parent as defined on  MSF \cite{rfc9033}. Consequently, if required, the number of added cells by BDPC in the reverse direction can be removed by MSF in the child-parent direction of the same link.

\begin{figure}
	\centering
	\includegraphics[scale=0.6]{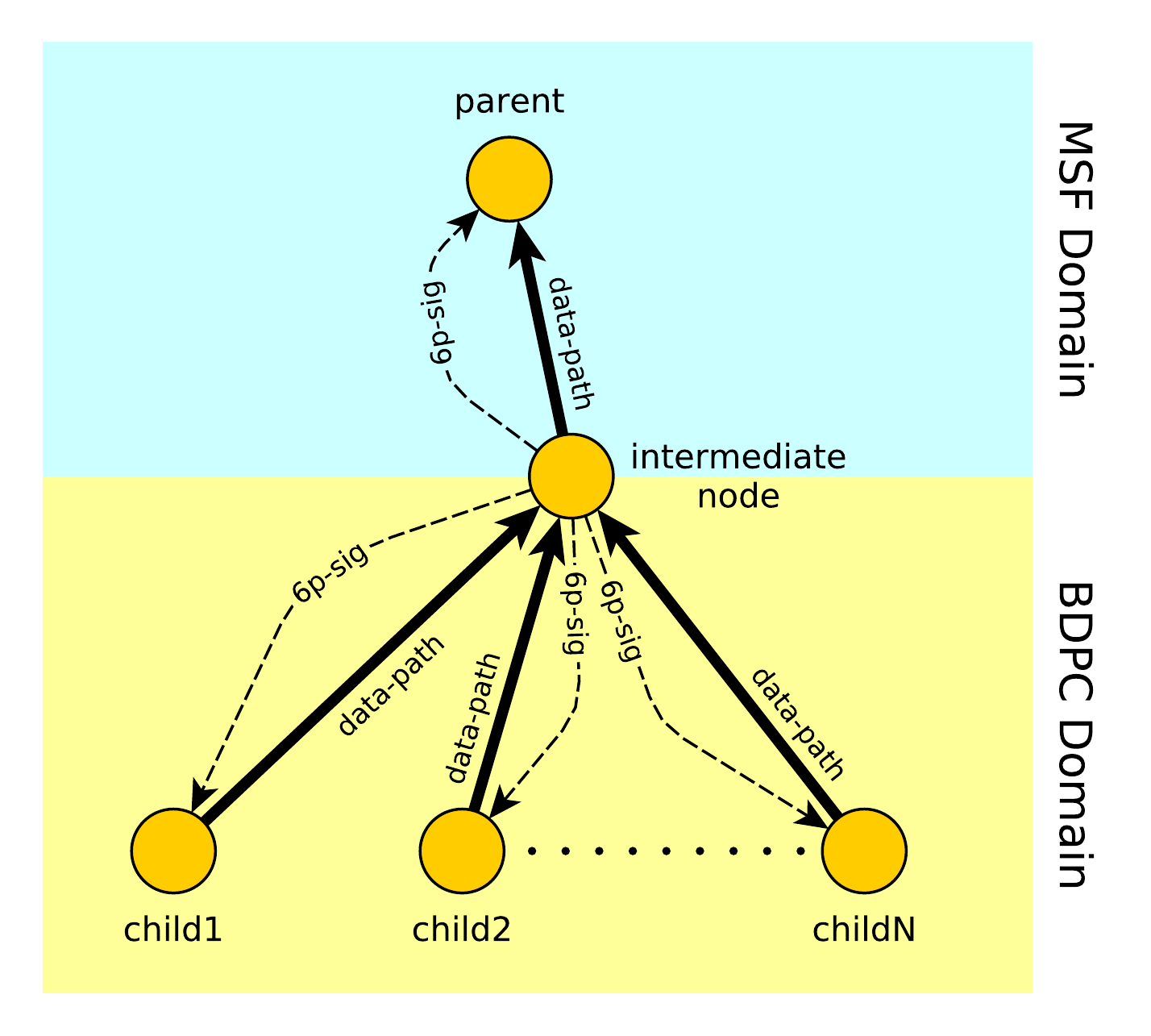}
	\caption{ 6P signaling  on the parent-child direction (Pre-Hop) controlled by algorithm \ref{alg:slots} of BDPC. MSF still controls cell assignment to the Next-Hop.}
	\label{fig:celdas_pre_hop}
\end{figure}

To summarize, the Scheduling Function integrates two complimentary behaviors: (i) the one defined by the standard, MSF, towards the preferred parent and (ii) the one performed by BDPC, towards the previous node. This is shown in Fig. \ref{fig_MSF_BDPC_stack} and Fig. \ref{fig:celdas_pre_hop}.

\section{Simulation and Results}
\label{section:results}

\subsection{Setup}

The 6TiSCH simulator \cite{municio_simulating_2019} includes the standard protocol implementations for MSF, 6P, and RPL in the 6TiSCH network stack. The RPL DIO message was modified to add the delay to root information to comply with the BDPC Scheduling Function (SF) definition.

We performed three experiments with 30 different seeds each. The first two experiments evaluate the performance of BDPC for different  values of \texttt{sfMax} and \texttt{sfMin}, while the remaining experiment is used as a reference, running the original MSF implementation.

The length of each simulation run is 10000 slotFrames (approximately 2.8 hours in simulation time) in order to allow the system stabilize after the initial network convergence stage. 

\begin{figure}[ht!]
    \centering
	\includegraphics[scale=0.63]{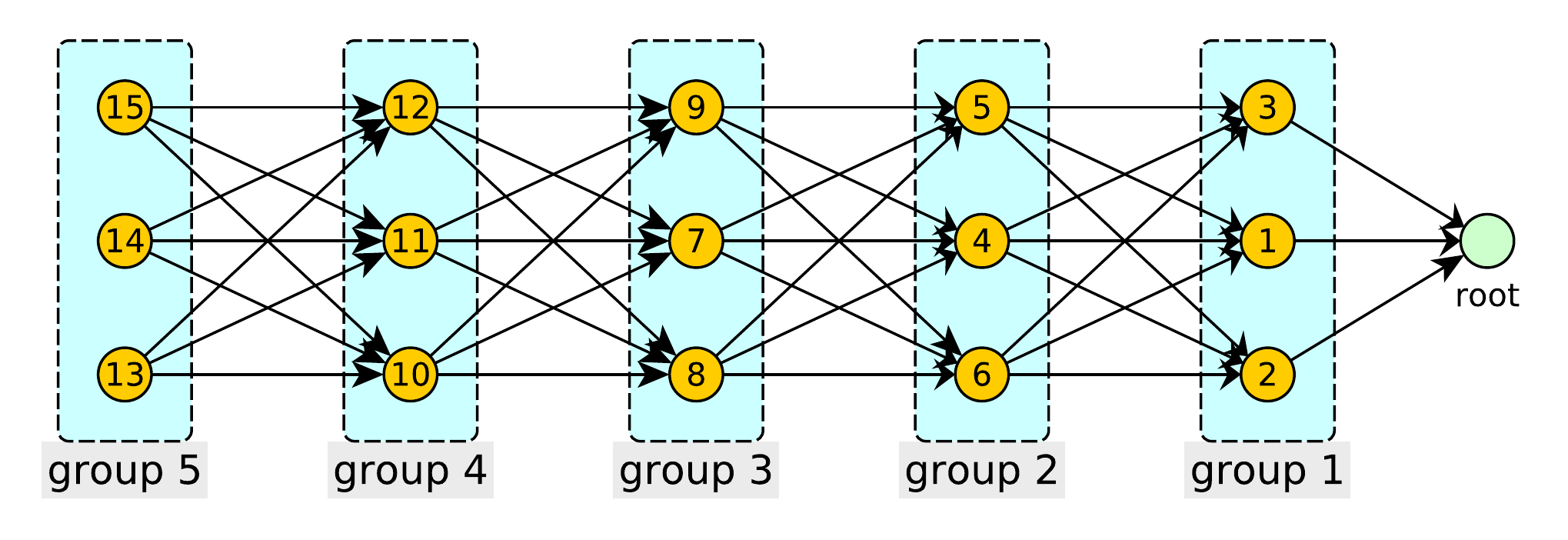}
	\caption{Simulation topology.  A 6TiSCH-based topology is expected to be multi-hop. The depth of the topology helps visualizing the effect of the delay on the data packets. }
	\label{fig:topology}
\end{figure}

The network topology is a hierarchy of groups of nodes (Fig. \ref{fig:topology}), each group containing three nodes. In the first group, each node has a link towards the destination (root) node, which is shown in green. In the rest of the groups, each node has  physical links towards three neighbors  in the group to the left, and three to the group on the right: nodes between even groups are not capable of establishing connections among them.  Consequently, there is no physical interference, for example, between group 3 and group 1, or between group 4 and 2, etc. 

The choice of the industrial IoT network topology is based both on technological constraints and typical use cases widely used in available literature.
A 6TiSCH-based network is typically multi-hop, as the one proposed in this paper. Several works such as \cite{ldsf,resf,llsf} use multi-hop topologies in their proposals for common industrial IoT 6TiSCH networks.  Moreover, multihop industrial topologies based on use cases are described in \cite{leonardi2018multi, leonardi2023mrt}.  

Also, a multi-hop topology allows for dynamic parent change instances during the simulation run. Parent change events are typical working conditions of route maintenance during the life of the wireless industrial IoT network and generate delay variations on the data packets traversing the network due to data packet buffering until the new route is established. 

The values of $\text{PDR}_{link} = 100\%$ and $\text{RSSI}_{link} = -10dB$ for each of the links, correspond to the default values provided by the simulator and they are generally used in other reference papers such as \cite{chang20206tisch,hauweele2020pushing, righetti2018analysis}.

All the nodes (except for the root) generate data traffic where the destination is the root node. This means that there are multiple data packet flows traversing the network in parallel, from the source nodes towards the destination (root) node. Each source node generates 90-Byte\footnote{The IEEE 802.15.4 header is 23 bytes long. This leaves room for a 104 bytes payload. A 90 bytes payload is a representative and avoids packet fragmentation.} data packets that are transmitted every 30s\footnote{The default period for packets in the simulator is 60s, but for the sake of creating a more realistic condition, the simulator is configured to use a 30s period between packets.}, with 0.05 variance\footnote{The variance of 0.05, equals to a standard deviation of 0.223s, which is the default value in the simulator. The rationale behind the variance value choice in the period between packets is that they should be transmitted within the first available transmission opportunity in the slotFrame, and that packet processing time may generate a variable delay on each of the intermedaite nodes. Moreover, data packets must arrive at the destination (root) node before the $deadline$, regardless of their transmission timestamp.}. The maximum delay threshold (\texttt{maxDelay}) is configured to 1.5s\footnote{In control networks, the dead time is the delay a signal takes from a controller output until it is measured and hence there is a response. The effect of dead time in a process needs to be systematically reduced because time delay makes the designed controller unstable \cite{sunori2017dead}. In \cite{lim1989generalized}, the delay ranges between 30s and 45s.}. The timeSlot duration is 10ms and the slotFrame is composed of 101 timeSlots, both default values extracted from the standard. Table \ref{tab:setup} summarizes the simulation parameters.

\begin{table}
\centering
\begin{tabular}{l|r}
\hline
\multicolumn{2}{c}{\textbf{General Parameters}}                                                        \\
\hline
Seeds             & 30 per experiment                                                                  \\
Nodes             & 15 + root                                                                          \\
Topology          & \begin{tabular}[c]{@{}r@{}}Multi-hop with parent\\ change possibility\end{tabular} \\
PDR$_{link}$         & 100.00\%                                                                           \\
RSSI$_{link}$        & -10dBm                                                                              \\
Simulation time   & 10000 slotFrames (2.8hs)                                                           \\
slotFrame         & 101 timeSlots                                                                      \\
timeSlot          & 10ms                                                                               \\
Channels          & 16                                                                                 \\
Packet Generation & 30s                                                                                \\
Packet Variance   & 0.05                                                                               \\
Packet Size       & 90 Bytes                                                                           \\
TSCH TX Queue Size     & 10 packets                                                                          \\
TSCH Max Retries     & 5                                                                     \\
Application \texttt{maxDelay}     & 1.5s                                                               \\
\hline
\multicolumn{2}{c}{\textbf{MSF}}                                                                       \\
\hline
\multicolumn{2}{c}{Original Implementation}     \\
\hline
\multicolumn{2}{c}{\textbf{BDPC}}                                                                      \\
\hline
BDPC\#1      & \begin{tabular}[c]{@{}r@{}}\texttt{sfMax} = 0.1\\ \texttt{sfMin} = 0.05\end{tabular}      \\
BDPC\#2       & \begin{tabular}[c]{@{}r@{}}\texttt{sfMax} = 0.0001\\ \texttt{sfMin} = 1e-05\end{tabular} \\
PreHop addCell    & 1                                                                                 \\
\hline
\end{tabular}
\caption{Simulation Setup parameters}
\label{tab:setup}
\end{table}

\subsection{Results}

\begin{figure}
\centering
	\includegraphics[scale=0.7]{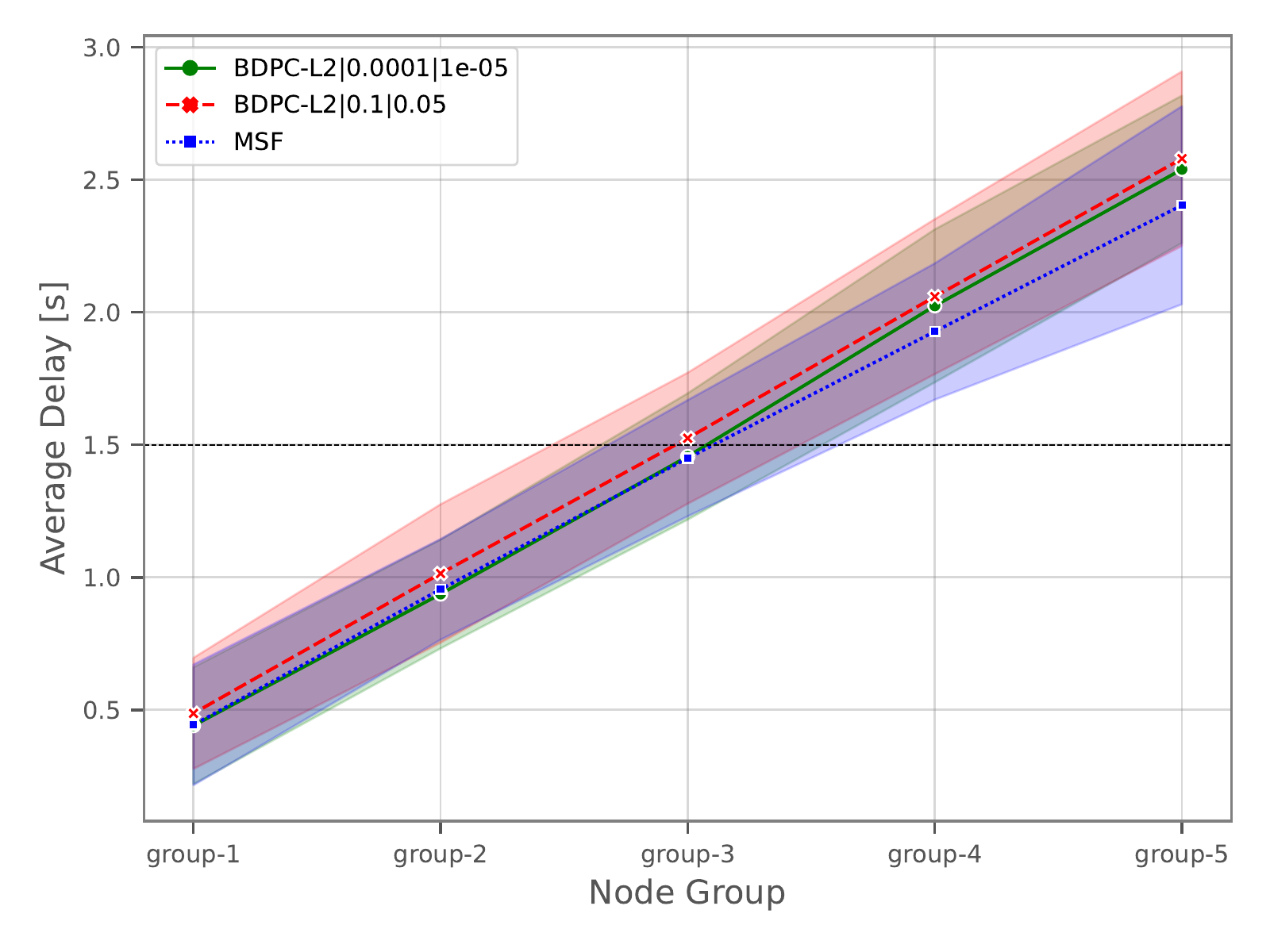}
	\caption{DIO Delay to destination (root) node measured at each node. The horizontal black-dashed line represents the maximum delay, \texttt{maxDelay}. The shaded area around the average delay to root, is the standard deviation for the series.}
	\label{fig:dio_delay2root_simul}
\end{figure}

Fig. \ref{fig:dio_delay2root_simul} shows the average delay observed by a node within its group, with respect to the destination (root) node, based on the information provided by the DIO messages. The horizontal dotted line shows the maximum delay (\texttt{maxDelay}). The delay to the destination (root) node, $d2r$, remains constant across the experiments because DIO messages use the Minimal-Cell, which belongs to the control plane, while BDPC manages the Data plane delay.

\begin{figure}
	\centering
	\includegraphics[scale=0.7]{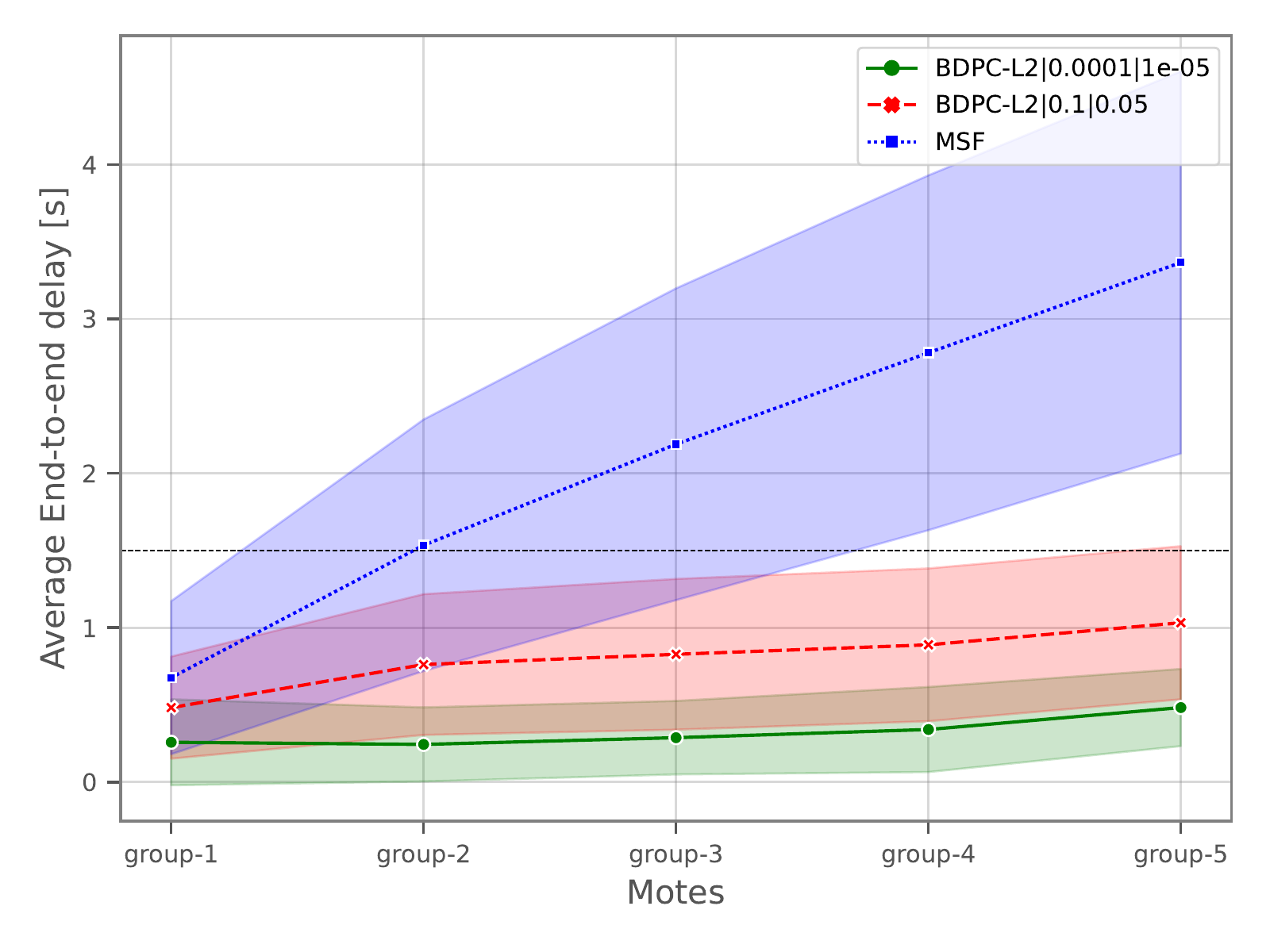}
	\caption{End-to-End Data packet delay measured at the destination (root) node. The horizontal black-dashed line represents the maximum delay, \texttt{maxDelay}. The shaded area around the average delay, is the standard deviation for the series.}
	\label{fig:data_delay2root_simul}
\end{figure}

The End-to-End Data packets delay measured at the destination (root) node is shown in Fig. \ref{fig:data_delay2root_simul}. Using MSF, the average delay exceeds the maximum allowed delay when the nodes reside on group-2 and backwards. When BDPC is enabled, the average delay stays below the maximum delay until the packets arrive at the destination (root) node. For example: the packets transmitted by a node within group-5, show an average delay measured at the destination (root) node of 1.03s and 0.48s for \texttt{sfMax} values of 0.1 and 0.0001, respectively.

\begin{figure}
	\centering
        \hspace*{-1.3cm}
	\includegraphics[scale=0.4]{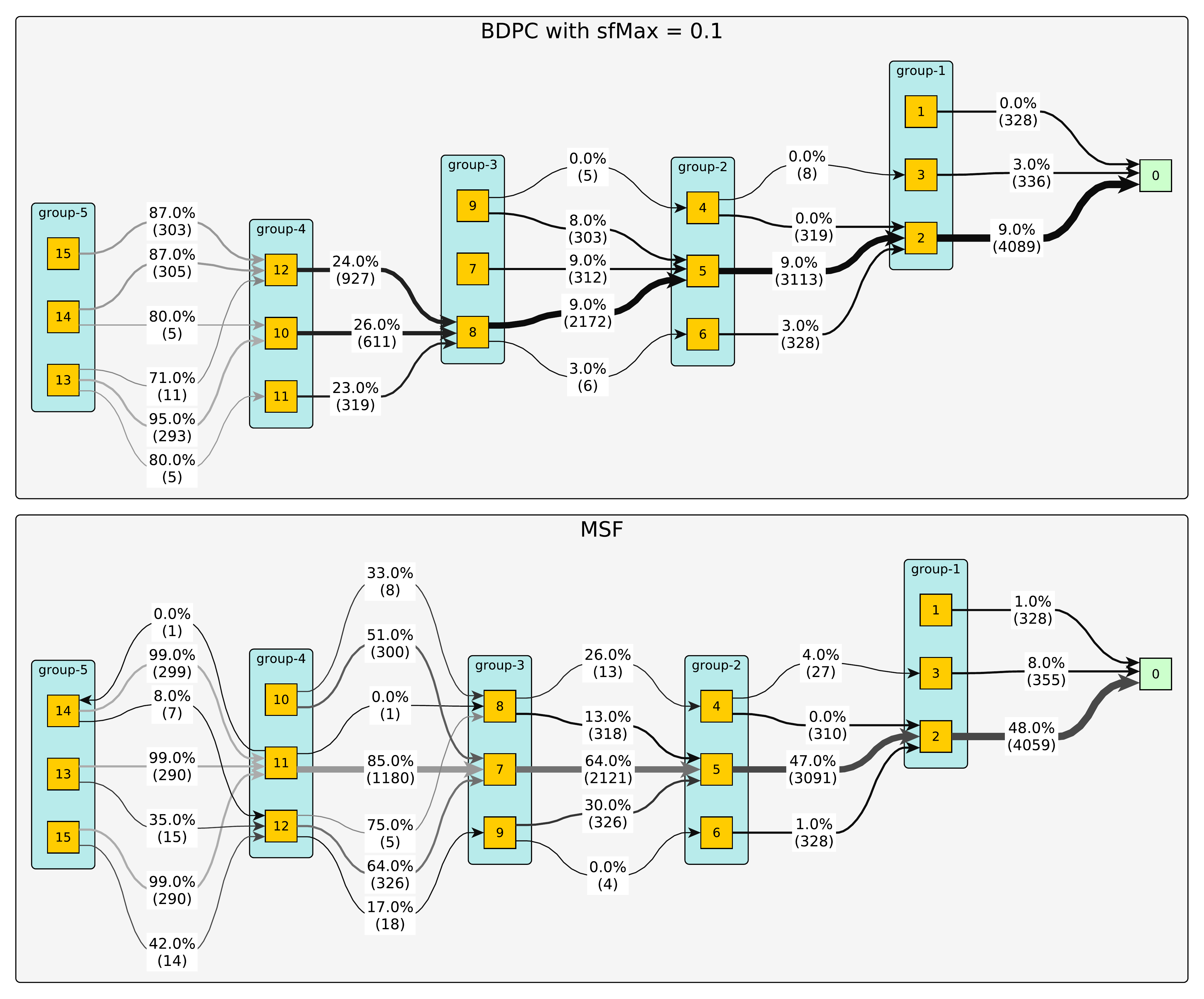}
	\caption{
	Example of the evolution of $latePaqs$ along the network. When MSF is used, $latePaqs$ is hardly reduced. On the other hand, when BDPC is used, the $latePaqs$ starts to decrease more aggressively because of the added cells in the parent-child direction. 
	}
	\label{fig:topology_latePaqs}
\end{figure}

Fig. \ref{fig:topology_latePaqs} shows an example of the different $latePaqs$ measurements that each intermediate node obtains from its children. When MSF is used, $latePaqs$ is hardly reduced and hence an important number of data packets arrive late at the destination (root) node. On the other hand, when BDPC is used, $latePaqs$ starts to decrease more aggressively because of the added cells in the parent-child direction.

For example, in Fig. \ref{fig:topology_latePaqs}, when using MSF, 48\% of the packets received by the destination (root) node from node 2, were late packets. Conversely, when using BDPC, only 9\% of the packets that the root node received from node 2, were late packets (which is coherent with the value of \texttt{sfMax}=10\%). It is important to note that the outgoing arrows mean aggregated traffic: the packets delivered by node 2 to the destination (root) node is a compound of its own packets plus the packets received from its child nodes. 

\begin{figure}
	\centering
	\includegraphics[scale=0.7]{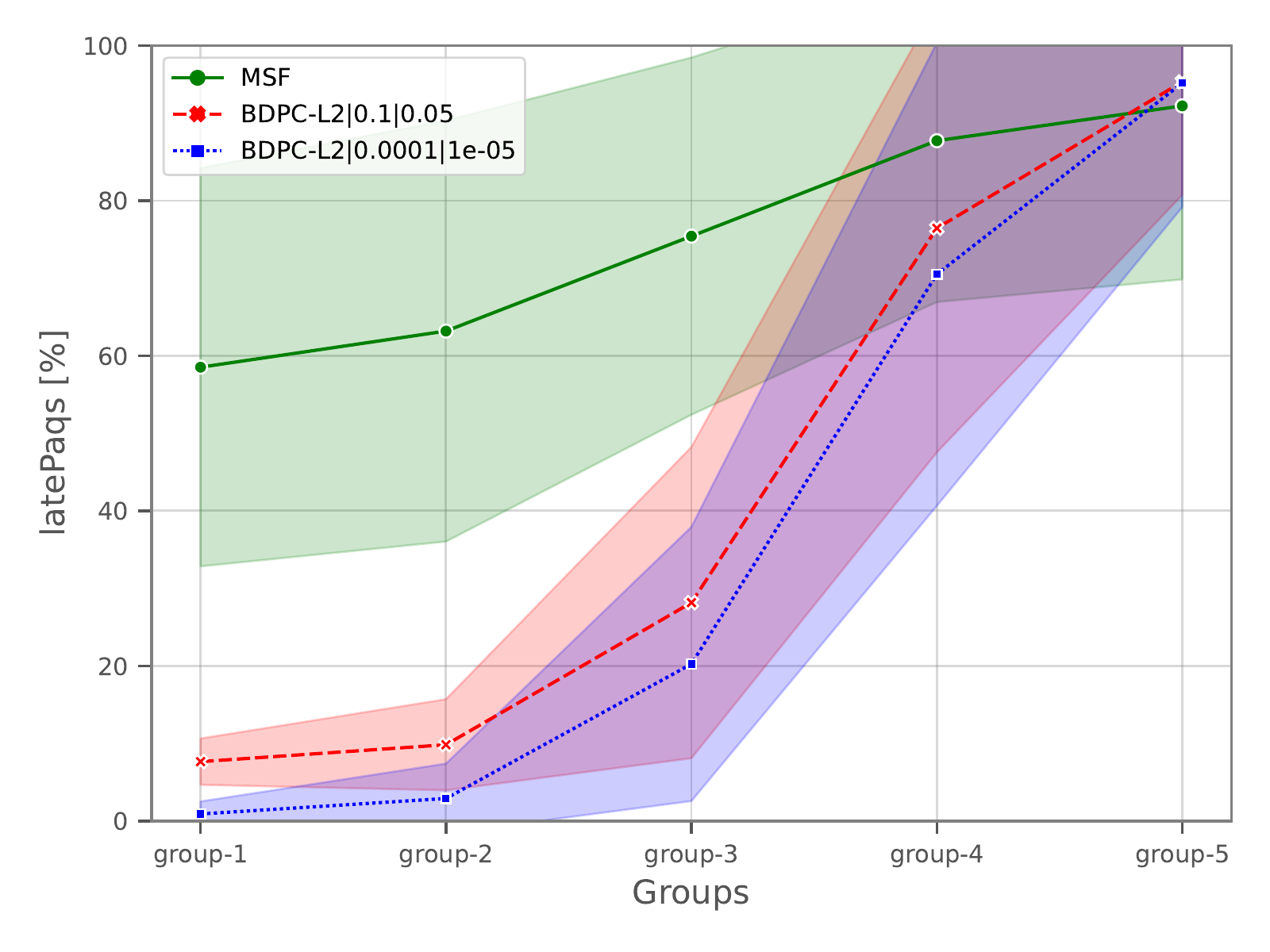}
	\caption{Evolution of $latePaqs$ along the network. The shaded area around the average $latePaqs$, is the standard deviation for the series. An example over the network topology can be seen in Fig. \ref{fig:topology_latePaqs}.}
	\label{fig:latePaqs}
\end{figure}

Fig. \ref{fig:latePaqs} shows the evolution of the variable $latePaqs$ along the network topology. It can be observed that $latePaqs$ value in groups away from the root is substantially higher compared to groups closer to the destination (root) node; this means that BDPC, reacting to $latePaqs$, adds \texttt{RX} cells to the PreHop recursively creating a chain that will reduce $latePaqs$ in nodes closer to the destination (root) node. A smaller $latePaqs$ means less packets arrive late at that intermediate node.

When the destination (root) node receives the packet, if it arrived before the $deadline$, that packet increments $inTime$ variable. Otherwise, the $delayed$ variable is incremented. The $lateAtRoot$ variable is then calculated as the ratio of delayed packets to the total amount of packets received by the destination (root) node as it can be observed in Eq. (\ref{eq_late_at_root}).

\begin{equation}
lateAtRoot_{IP} = f(deadline) =  \frac{delayed}{delayed+inTime}
\label{eq_late_at_root}
\end{equation}

\bigskip

 The source node is identified by the IP address of the incoming packets and hence, there is one different $lateAtRoot$ value for each source node IP in the network.

Since the packets arriving at the destination (root) will not be forwarded to another hop in the path, they are counted to calculate $PDR_{e2e}$. $PDR_{e2e}$ is the ratio of received packets to the number of transmitted packets, at the root node, i.e. $PDR_{e2e}= n_{rx}/n_{tx}$. The ratio of those packets arriving late at the destination (root) node is measured by the variable $lateAtRoot$. Consequently, the $PDR_{e2e}$ value before the $deadline$ is:

\begin{equation}
	\begin{split}
		PDR_{e2e | d \leq deadline} & = 1 - lateAtRoot \\
		p(d \leq deadline) & =  1 - lateAtRoot \\
	\end{split}
	\label{eq_chap3_one_minus_latePaqs_e2e}
\end{equation}

\bigskip

If $lateAtRoot$ is replaced by \texttt{sfMax} in Eq. (\ref{eq_chap3_one_minus_latePaqs_e2e}), then the probability of the end-to-end delay when the arrival time is lower or equal to the maximum deadline \texttt{maxDelay} is equal to $(1-\texttt{sfMax})$ as it can be observed in Eq. (\ref{eq:one-sfMax}).

\begin{equation}
p(d \leq \texttt{maxDelay}) = 1-\texttt{sfMax}
\label{eq:one-sfMax}
\end{equation}

\bigskip

Hence, $(1-\texttt{sfMax})$ is the probability that data packets traversing the network towards the destination (root) node will arrive before the deadline.

\begin{figure}
	\centering
	\includegraphics[scale=0.7]{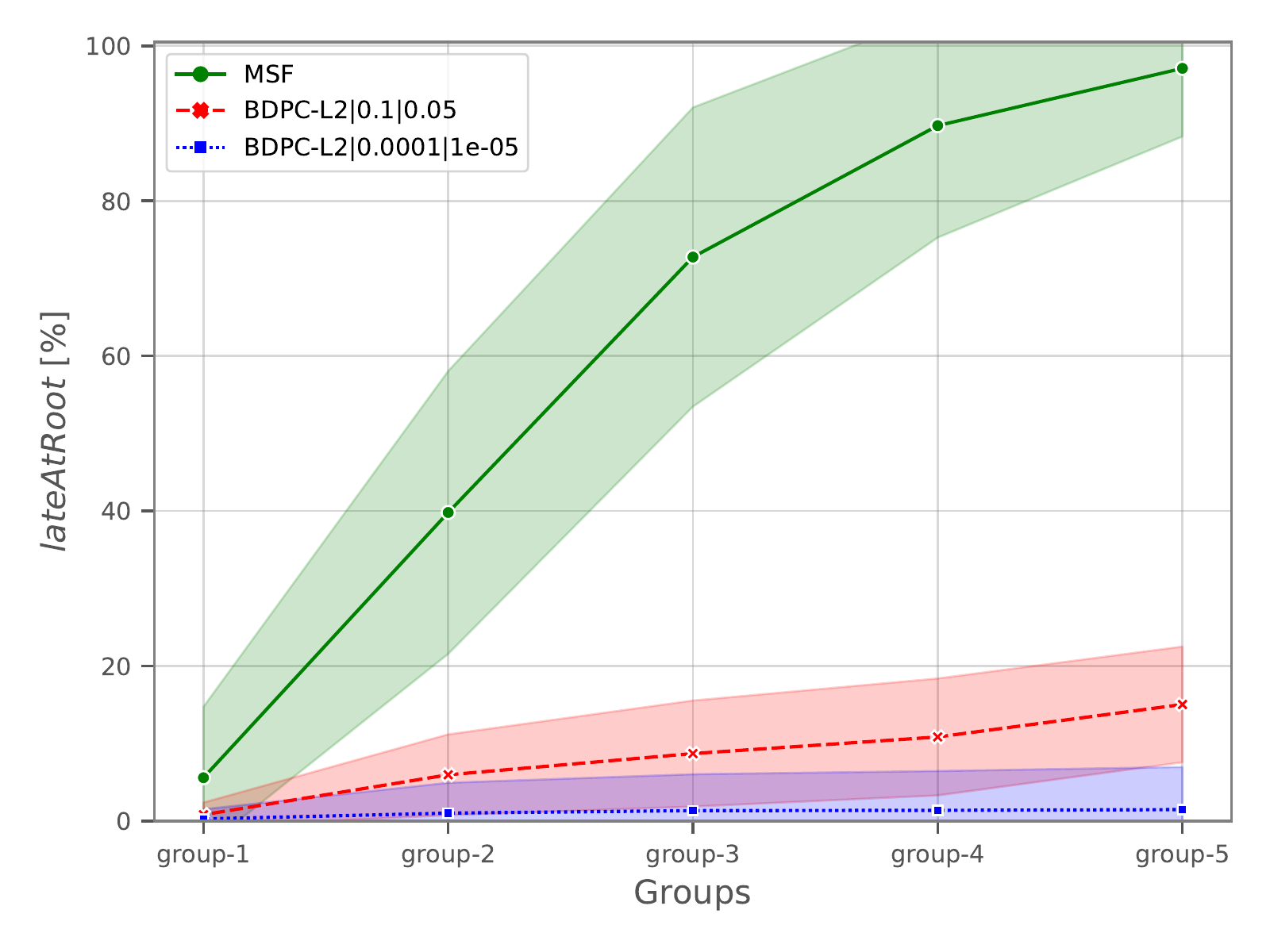}
	\caption{Data packets arriving late at the root node. Packets are originated in each node within each group on the network. The shaded area around the average late packets arrival ratio is the standard deviation for the series.}
	\label{fig:data_packets_late}
\end{figure}

Fig. \ref{fig:data_packets_late} describes the number of late packets arriving to the root from each source group. As such, 40\% of the packets which were generated from group 2 arrived after the deadline when using MSF as the Scheduling Function. On the other hand, when using BDPC, this ratio drops down to less than 6\% for packets generated from the same group. Thus, with BDPC, the smaller \texttt{sfMax} is, the lower will be the percentage of packets arriving late at the destination. For example, for data packets leaving a node within group-5, the average ratio of packets arriving late at the destination (root) node is 97.34\% when using MSF; when using BDPC this ratio drops to 14.06\% and 0.29\% for \texttt{sfMax} values of 0.1 and 0.0001, respectively.

By comparing the results of Fig. \ref{fig:latePaqs} and Fig. \ref{fig:data_packets_late}, i.e., $latePaqs$ versus $lateAtRoot$, it can be observed that even though $latePaqs$ is high in groups away from the destination (root) node, the number of packets originated in such groups (i.e., group-5) that actually arrive late when using BDPC, is less than 20\% while for the case of MSF is close to 100\% of late packets. Since $latePaqs$ measures the ratio of packets arriving late at an intermediate node, the lower $latePaqs$ becomes means less late packets arriving late at their destination. As a consequence, BDPC has the objective of reducing $latePaqs$ by using algorithm \ref{alg:slots} and thus adding \texttt{RX} cells to the PreHop. This results in a $latePaqs$ decrease in nodes closer to the destination (root) node and, as a consequence, a decrease in the actual ratio of packets arriving late at their destination. Table \ref{tab:data_packets_late_vs_latePaqs}, shows in detail the  ratio of packets arriving late at the root node  versus its corresponding value of $latePaqs$ for data packets originated in groups for different number of hops away from the destination (root) node.

\begin{table}
\centering
    \begin{tabular}{l|crr}
    \hline
             run &    group &  $latePaqs$ \%& $lateAtRoot$ \%\\
\hline
MSF & group-1 &   58.52262 &  6.33998 \\
MSF & group-5 &   92.22803 & 97.34692 \\
BDPC-L2|\texttt{sfMax}=0.0001 & group-1 &    0.89990 &  0.12184 \\
BDPC-L2|\texttt{sfMax}=0.0001 & group-5 &   95.17487 &  0.29296 \\
BDPC-L2|\texttt{sfMax}=0.1 & group-1 &    7.64576 &  0.71938 \\
BDPC-L2|\texttt{sfMax}=0.1 & group-5 &   95.33046 & 14.06473 \\

    \hline
    \end{tabular}
\caption{Ratio of data packets arriving late at the destination (root) node versus $latePaqs$ at intermediate nodes in groups 1 and 5. When using BDPC, an increment of $latePaqs$ does not provoke and increment on the ratio of late packets at the destination (root) node.}    
\label{tab:data_packets_late_vs_latePaqs}
\end{table}

\begin{figure}
	\centering
	\includegraphics[scale=.37]{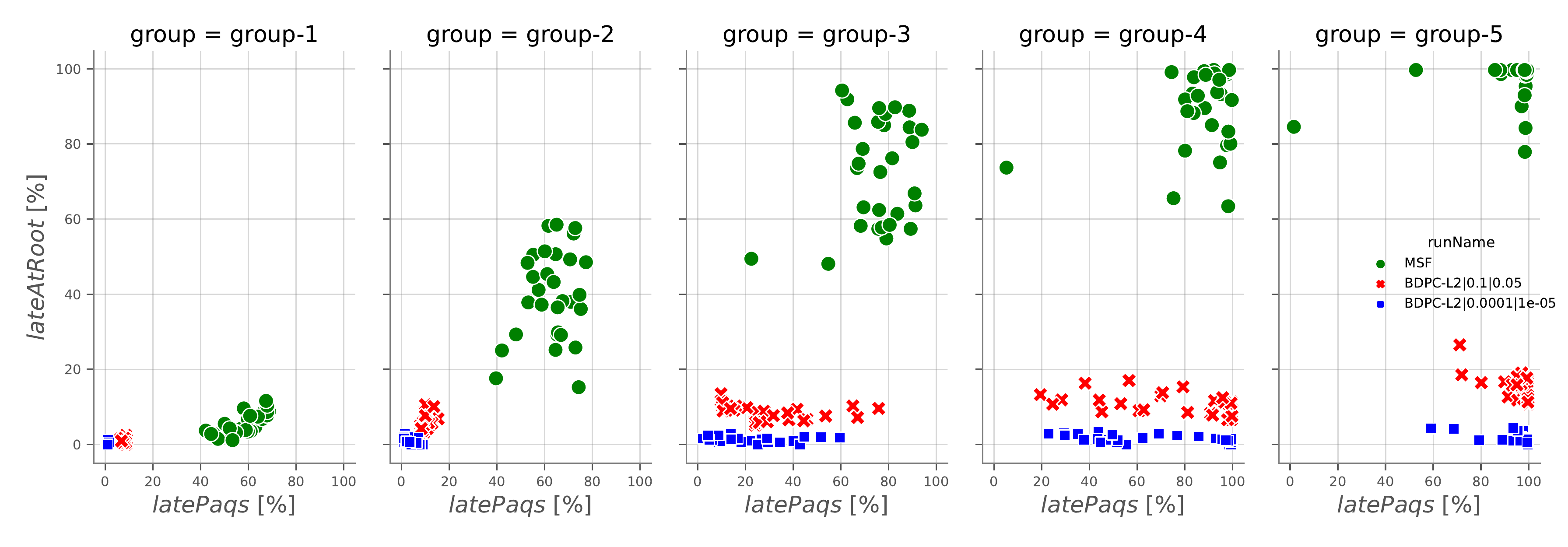}
	\caption{Evolution of $latePaqs$ along the network. The y-axis shows the value of $lateAtRoot$, which represents the overall late packets arriving at the root node from each group in the network. Table \ref{tab:data_packets_late_vs_latePaqs} summarizes the graphics' average values for groups 1 and 5}
	\label{fig:data_packets_late_vs_latePaqs}
\end{figure}

Fig. \ref{fig:data_packets_late_vs_latePaqs} shows a different analysis from a combination of what can be observed on Fig. \ref{fig:latePaqs} and Fig. \ref{fig:data_packets_late}. Fig. \ref{fig:data_packets_late_vs_latePaqs} is a scatter plot that depicts the distribution of the percentage of late packets at the root node for the full range of values of $latePaqs$ along the different groups of the topology. It can be observed that $latePaqs$ has a different distribution for each group: the measurements of $latePaqs$ are concentrated to the right for groups away from the root node (i.e., group-5). However, an increase of $latePaqs$ rate using BDPC does not follow an increase of the number of data packets arriving late at the root node. This is not the case for MSF, where an increment of $latePaqs$ results in an increment of the rate of late packets at the root node. The average packet delivery rate values for the three configurations can be observed in Table \ref{tab:data_packets_late_vs_latePaqs}.

\begin{figure}
	\centering
	
	\subfloat[Histogram of the E2E Delay]{ \label{fig:e2e_delay_hist}\includegraphics   [width=.48\textwidth]{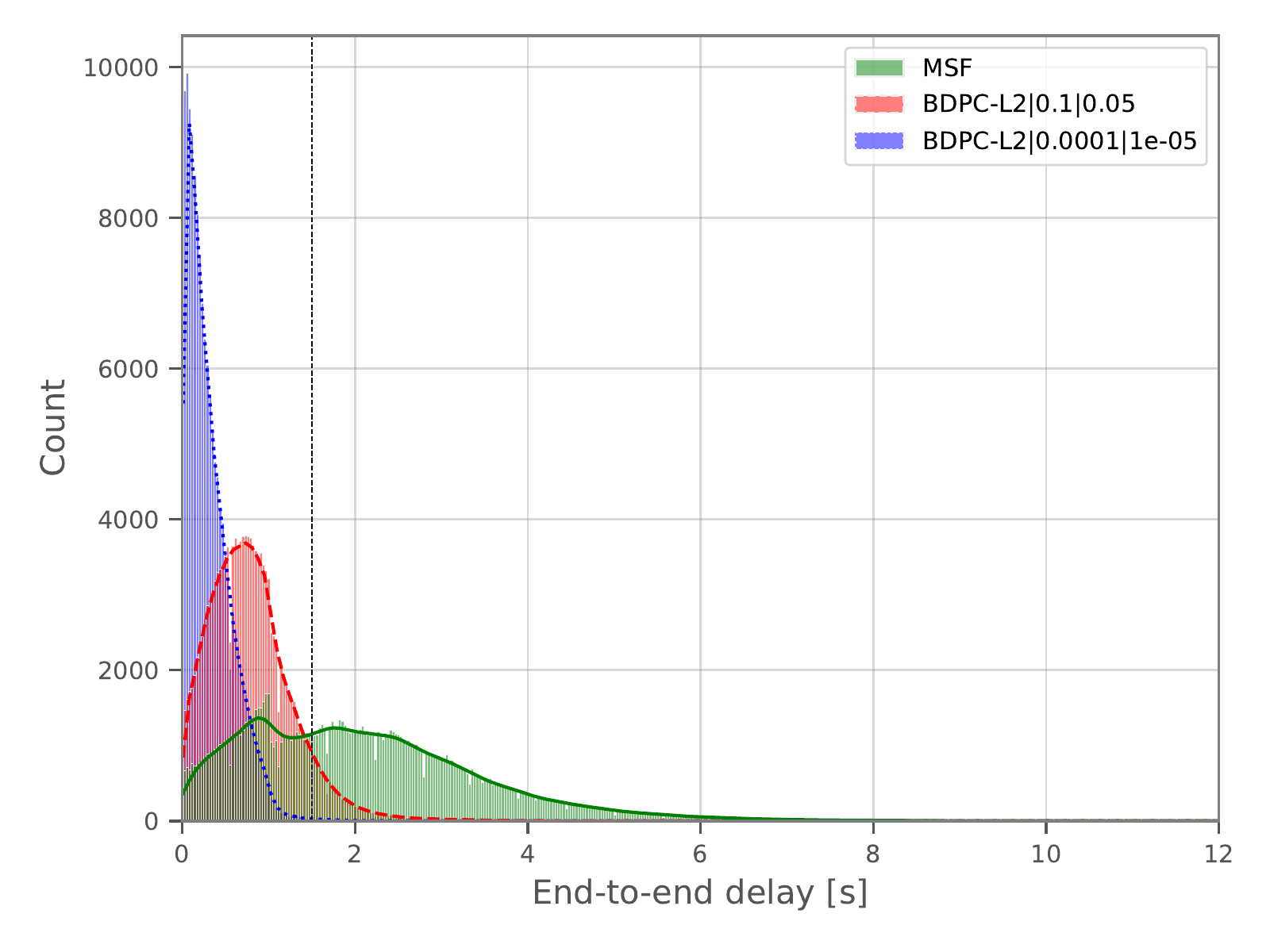}} \hfill
	\subfloat[Cumulative Distribution of the E2E Delay]{\label{fig:e2e_delay_cdf}\includegraphics [width=.48\textwidth]{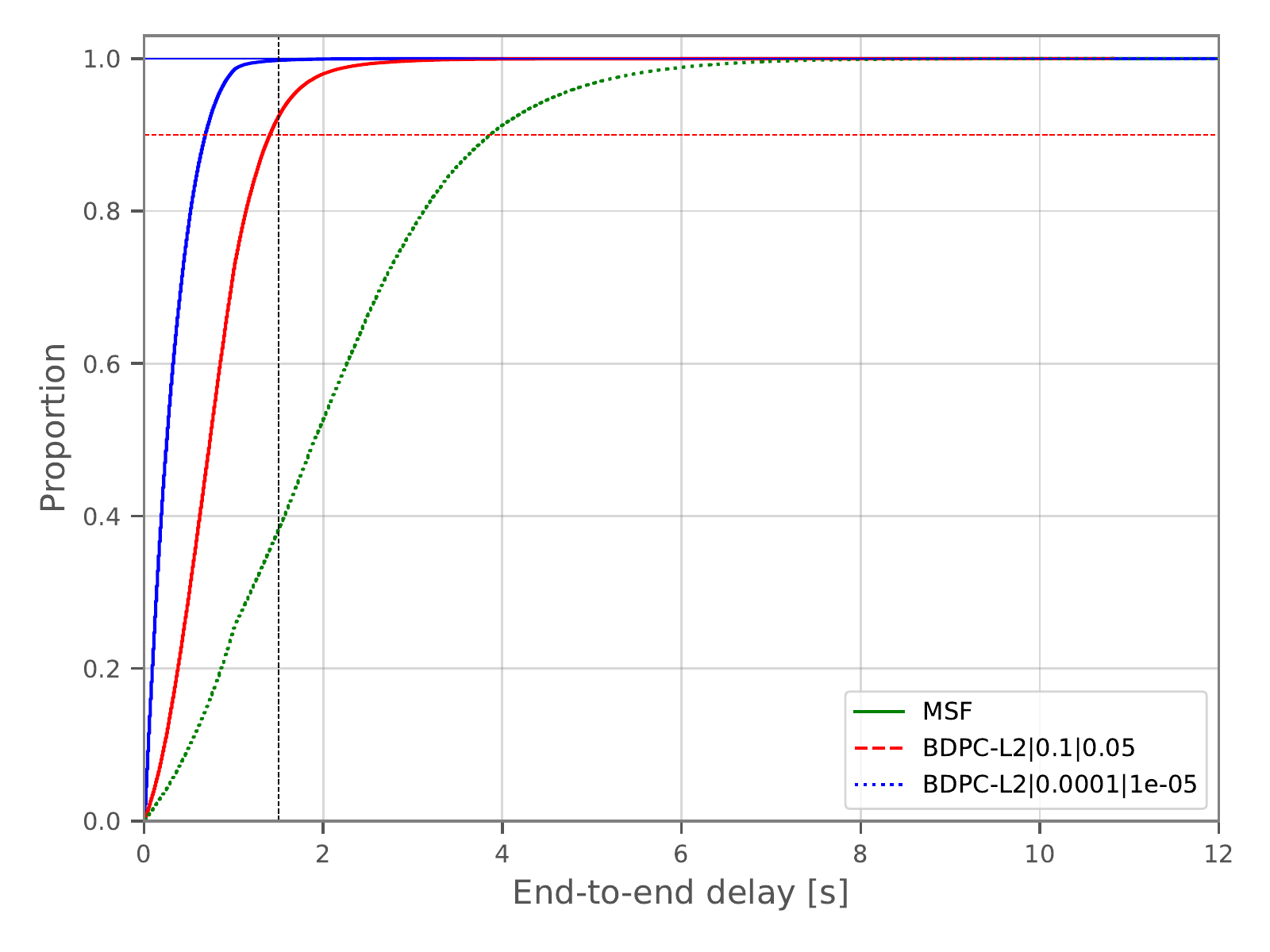}} \hfill

	\caption{E2E Delay, both histogram and cumulative distributions. The vertical dashed line represents the maximum delay, \texttt{maxDelay}, tolerated by the application. When BDPC is used, the $PDR$ under the deadline is equal to $1-\texttt{sfMax}$.}
	\label{fig:data_e2e_delay}
\end{figure}

The overall E2E delay performance of the network can be observed in Fig. \ref{fig:data_e2e_delay}. The vertical dotted line shows the maximum delay defined by the application. In the cumulative plot, the horizontal dotted lines represent the value of $(1-\texttt{sfMax})$. According to Eq. (\ref{eq:one-sfMax}), this value represents the overall rate of packets that will arrive at the destination within the deadline.

Moreover, MSF performance shows that only 38.5\% of the data packets will arrive within the deadline, while BDPC guarantees that the packets arriving in term will be 92.5\% and 99.8\% for \texttt{sfMax} values of 0.1 and 0.0001, respectively. If MSF is expected to achieve a packet delivery (before deadline) rate of 90\%, the maximum delay tolerated by the application must be 4s or longer. Consequently, not only BDPC enables a better deadline compliance, but also the \texttt{sfMax} parameter can finely control the overall rate of packets that will arrive within the deadline.


\begin{figure}
	\centering
	
	\subfloat[$PDR_{e2e}$ at the root node when using MSF alone]{ \label{fig:lostNetwork}\includegraphics   [width=.45\textwidth]{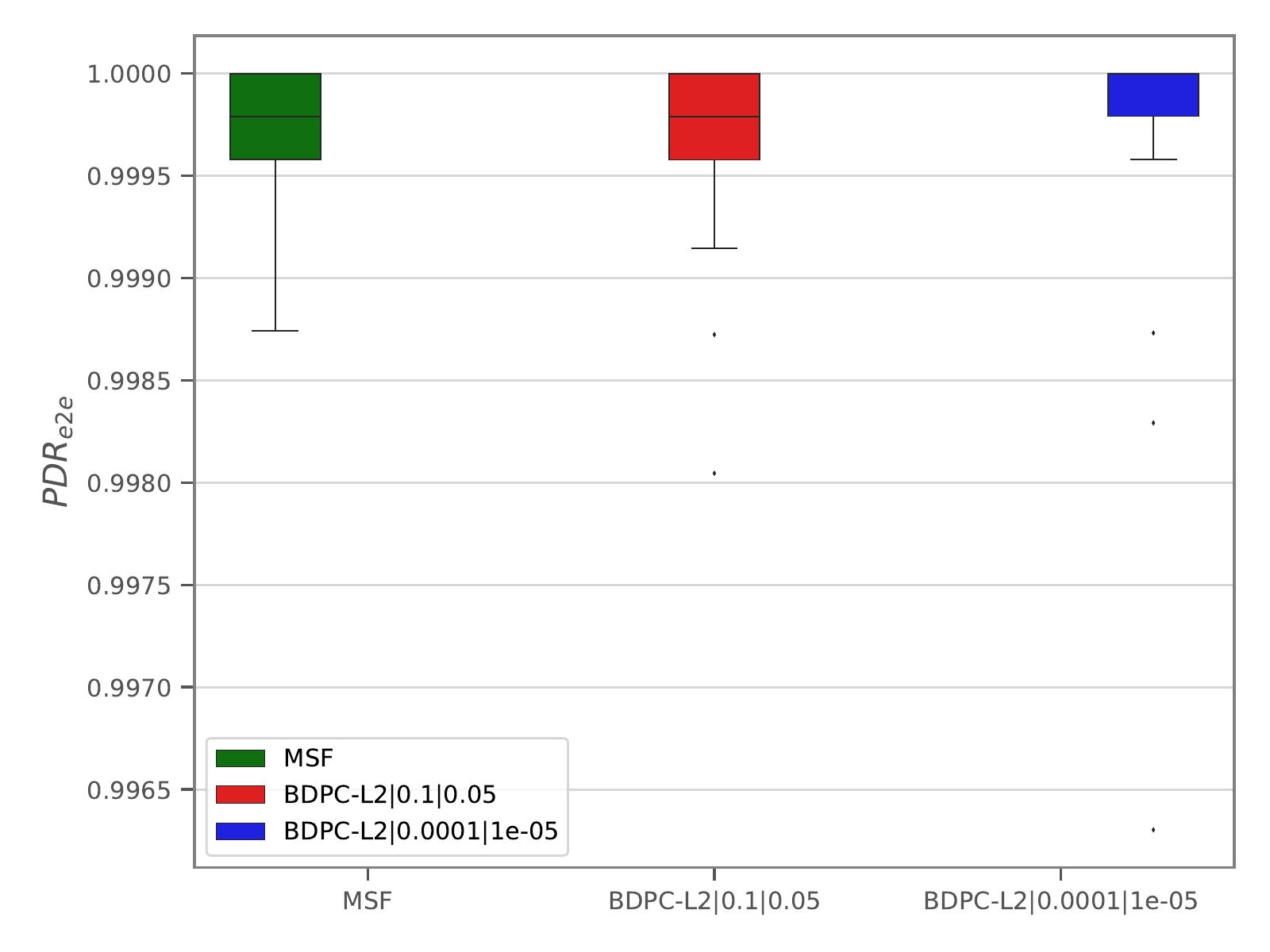}} \hfill
	\subfloat[$PDR_{e2e}$ at the root node, considering \texttt{maxDelay}]{\label{fig:lostDelay}\includegraphics [width=.45\textwidth]{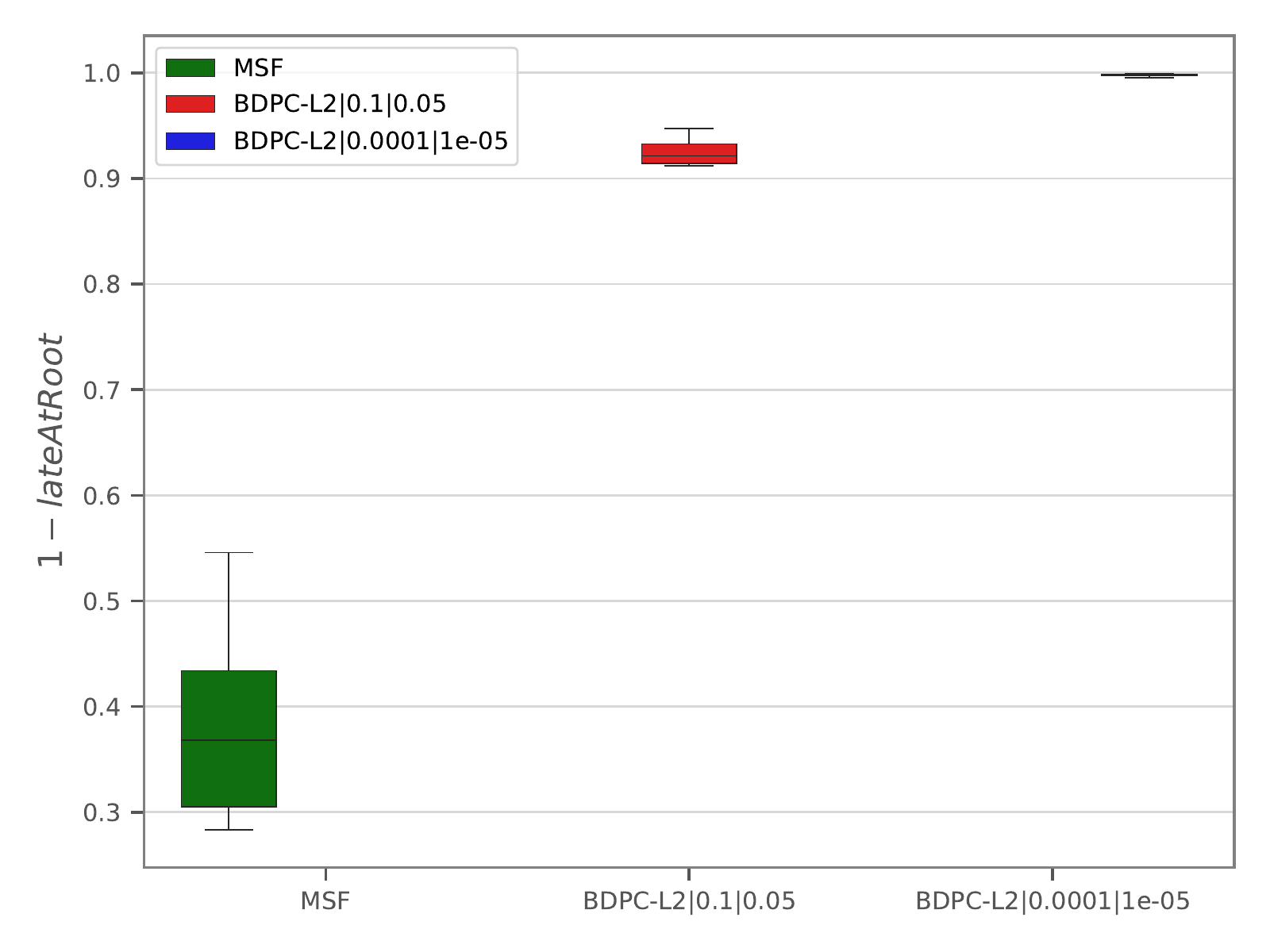}} \hfill

	\caption{Comparison of $PDR_{e2e}$ when using MSF versus when using BDPC, considering \texttt{maxDelay}}
	\label{fig_chap5_joint_msf_vs_all_pkPeriod}
\end{figure}	

Fig. \ref{fig:lostNetwork} and Fig. \ref{fig:lostDelay} show the packet delivery rate ($PDR_{e2e}$) at the application level. The first figure shows the rate of packets that successfully arrived at the root node, for each of the series. Since the $PDR_{link}$ is configured to 100\%, the $PDR_{e2e}$ at the application level is close to 99.975\%. However, when the packet deadline is used as a criteria to discard packets, as it is in realtime industrial IoT networks, Fig. \ref{fig:lostDelay} shows that BDPC guarantees the arrival of more than 90\% of the packets, compared to MSF which guarantees less than 40\% of the packets arriving on time. Table \ref{tab:arrival_rate} summarizes the graphics' values as a reference.

\begin{figure}
	\centering
	\includegraphics[scale=0.7]{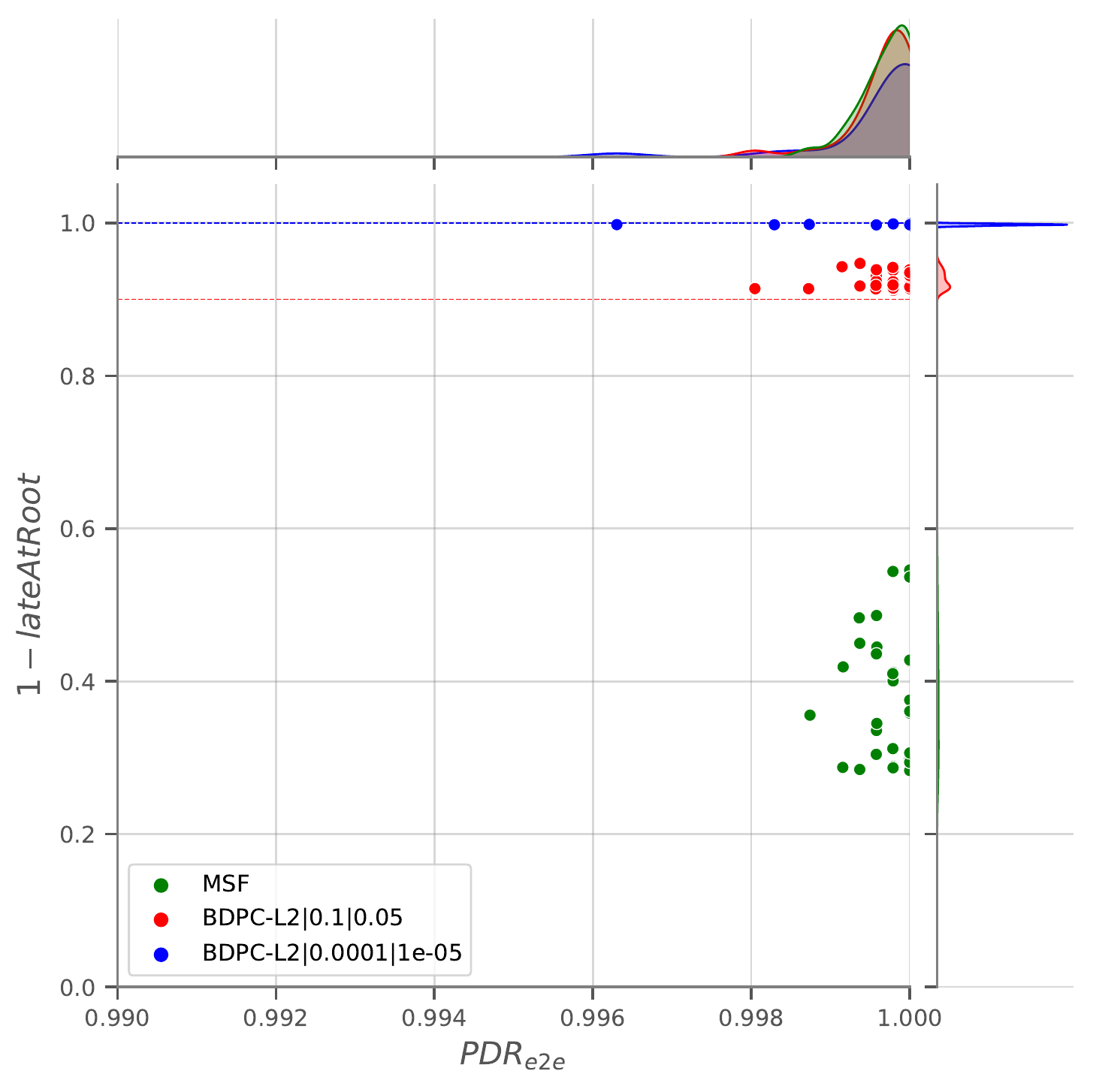}
	\caption{Joint Plot of $PDR_{e2e}$ vs $1-lateAtRoot$. The horizontal red and blue dashed lines, represent the values of $(1-\texttt{sfMax})$, for \texttt{sfMax} values of 0.1 and 0.0001 respectively. Both MSF and BDPC guarantee a high $PDR_{e2e}$, but only when BDPC is used, packets arrive before the maximum deadline. Table \ref{tab:arrival_rate} summarizes the graphics' average $PDR_{e2e}$ values as a reference}
	\label{fig:joint}
\end{figure}

Fig. \ref{fig:joint} correlates the results of Fig. \ref{fig:lostNetwork} and \ref{fig:lostDelay}. The horizontal red and green dashed lines, represent the values of $(1-\texttt{sfMax})$, for \texttt{sfMax} values of 0.1 and 0.0001 respectively. In this figure it can be observed that even though all the alternatives deliver a $PDR_{e2e}$ above 99.8\%, only when using BDPC the packets arrive before the deadline. Furthermore, the smaller \texttt{sfMax} is, the lesser jitter can be obtained for the packets arriving before the deadline.

\begin{table}
\centering
 \begin{tabular}{l|cc}
    \hline
             run &  $1-lateAtRoot$ &  $PDR_{e2e}$ \\
\hline
BDPC-L2|\texttt{sfMax}=0.0001 & 0.99766 & 0.99972 \\
    BDPC-L2|\texttt{sfMax}=0.1 & 0.92459 & 0.99968 \\
                 MSF & 0.38329 & 0.99971 \\

    \hline
    \end{tabular}
\caption{Average packet delivery rate when the \texttt{maxDelay} limit is enabled}    
\label{tab:arrival_rate}
\end{table}

\begin{figure}
	\centering
	\includegraphics[scale=0.7]{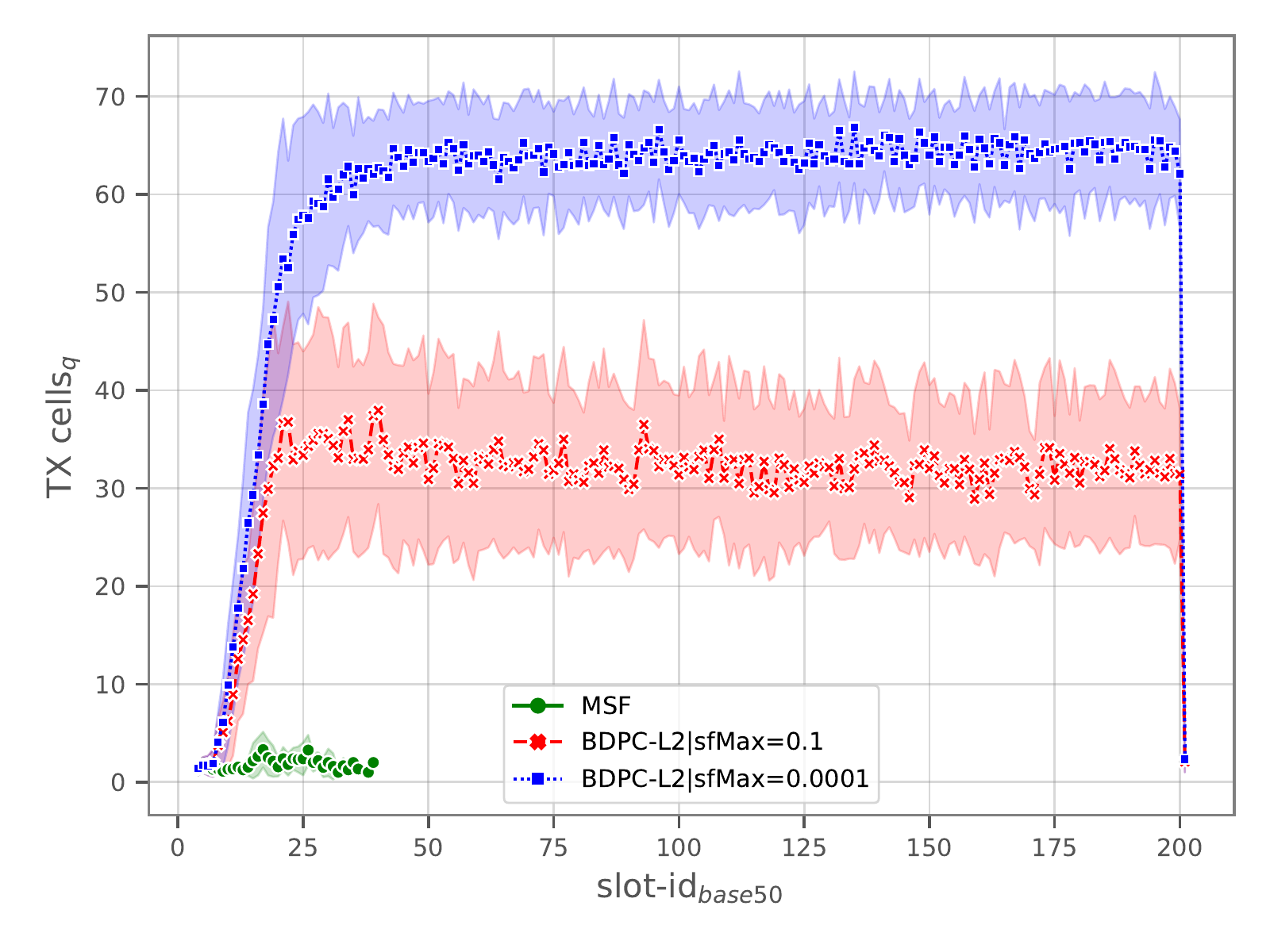}
	\caption{
	Reserved dedicated \texttt{TX} cells along the simulation run. The count of \texttt{TX} used cells are grouped over the span of 50 slotFrames (hence $50 \times 200=10000$ slotFrames, which is the duration of the simulation run). The shaded area is the standard deviation of the series. After slotFrame 2000, MSF stops allocating resources, because traffic load is already stable. However, BDPC keeps on sensing the data traffic value every time a packet traverses a node and adjusts cell capacity to cope with the application requirements.
	}
	\label{fig:txCellsTS}
\end{figure}

But yet, why does it work? The previous results have presented the performance in terms of $PDR_{e2e}$ vs $1-lateAtRoot$ (Fig. \ref{fig:joint}), end-to-end delay distribution (Fig. \ref{fig:data_e2e_delay}) and the distribution of the $latePaqs$ variable versus the $lateAtRoot$ variable (Fig. \ref{fig:data_packets_late_vs_latePaqs}). Fig. \ref{fig:txCellsTS} shows the dedicated reserved \texttt{TX} cells along the simulation run. For the case of MSF, after the 2000th slotFrame, the scheduling function has already clustered all the necessary cells for the current traffic load. On the other hand, BDPC senses the traffic value every time a packet traverses a node, and precisely adjusts the required cells to agree to the application deadline requirements 
Once the traffic has converged, BDPC only triggers small changes to the schedule, by means of algorithm \ref{alg:slots}. In general, BDPC with sfMax=0.1 requires less \texttt{TX} cells to cope with the application requirements, while BDPC with sfMax=0.0001 requires the double of capacity, on average, for the same purpose.

\begin{figure}
	\centering
	\includegraphics[scale=0.65]{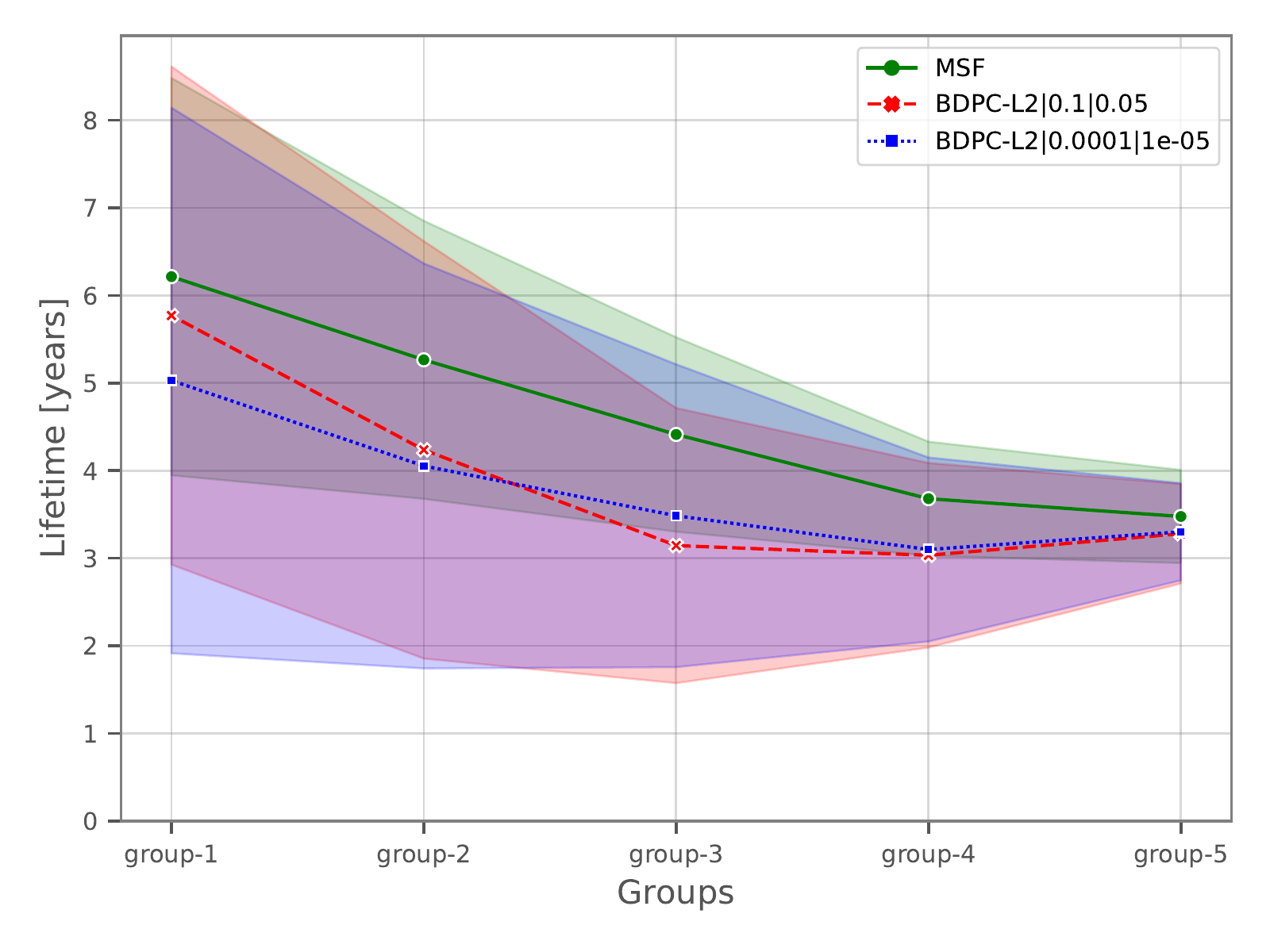}
	\caption{Node lifetime. The lifetime of the network is defined by the period until at least one of the nodes in the network depletes its energy storage and stops responding.}
	\label{fig:mote_lifetime}
\end{figure}

Finally, Fig. \ref{fig:mote_lifetime} shows the expected lifetime for each group in the network, for each of the experiments\footnote{To calculate the node lifetime, the charge of a 2821.5mAh AA battery is considered as a reference. The energy consumption is calculated by counting the number of TX/RX/ACK/Listen/Sleep radio operations performed by a node.}.  The life of the network is defined by the longest time all nodes are still alive, which corresponds to the node with the shortest lifetime. From this point of view, all experiments behave similarly\footnote{Nodes closer to the root have lower power consumption than nodes located farther away is counter-intuitive; this is a result of the network cell allocation process using management frames. Nodes closer to the root need less packet management exchanges given that the traffic is already clustered, while power consumption at the extremes requires more negotiation instances.}. Consequently, the average lifetime of the network using BDPC is 3.06 and 3.11 years, for the experiments with \texttt{sfMax} values of 0.1 and 0.0001 respectively, while the network would run 3.48 years with MSF; a lifetime increase of only 4.4 months. By sacrificing 11.8\% of the network lifetime when using MSF, 99.8\% of the packets could reach their destination within the deadline.

\section{Conclusion}
\label{section:conclusion}

This paper proposes a new approach to achieve realtime capabilities in wireless industrial IoT control systems. In the industrial context, two critical factors define performance: an extremely low packet loss rate and deadline compliance at application data packets. Packets arriving late are discarded at the application layer guaranteeing the delivery of data packets within a maximum and predictable deadline is of utmost importance. Moreover, it is not only important to guarantee an extremely high $PDR_{e2e}$ as a general case, it is also important to guarantee an extremely high $PDR_{e2e}$ with realtime compliant deadlines.

We proposed BDPC (Bounded Delay Packet Control) to tackle this problem. BDPC is a mechanism that combines the knowledge of the maximum delay tolerated by the application with the observed delay towards the root node to allocate network resources. Additionally, we created the $latePaqs$ metric. $latePaqs$ informs the intermediate node about the rate of late packets traversing it. We also defined two thresholds called \texttt{sfMax} and \texttt{sfMin} to establish the number of packets allowed to arrive later than the deadline, and finally, we designed a new allocation request mechanism in the inverse, parent-child direction (PreHop). Furhtermore, BDPC is agnostic to traffic pattern, flow isolation or topology structure.

BDPC shows a substantial improvement over the standard behavior of MSF, where the network delivers 99.8\% of the packets within the deadline, for a value of $\texttt{sfMax}=0.0001$, compared to the standard MSF, which guarantees on-time delivery of only 38.5\% of the packets within the deadline, for the same network and application with only a network lifetime decrease of 11.8\% (equivalent to 4.4 months).  

Finally, this proposal has the potential to be implemented in the current standard, given that the solution only involves minor modifications to the 6TiSCH protocol stack.
\section*{Acknowledgment}
The authors would like to thank the CYTED AgIoT Project (520rt0011), CORFO CoTH2O ``Consorcio de Gestión de Recursos Hídricos para la Macrozona Centro-Sur'', Proyecto Asociativo UDP ``Plataformas Digitales como modelo
organizacional'' and 19STIC-09 ``WirelessWine'' STIC-AmSud, for their support during the development of this research work. 


 \bibliographystyle{elsarticle-num} 
 \bibliography{cas-refs}

\begin{thebibliography}{10}
\expandafter\ifx\csname url\endcsname\relax
  \def\url#1{\texttt{#1}}\fi
\expandafter\ifx\csname urlprefix\endcsname\relax\def\urlprefix{URL }\fi
\expandafter\ifx\csname href\endcsname\relax
  \def\href#1#2{#2} \def\path#1{#1}\fi

\bibitem{ryu2005urgency}
S.~Ryu, B.~Ryu, H.~Seo, M.~Shin, Urgency and efficiency based packet scheduling
  algorithm for ofdma wireless system, in: IEEE International Conference on
  Communications, 2005. ICC 2005. 2005, Vol.~4, IEEE, 2005, pp. 2779--2785.

\bibitem{sunori2017dead}
S.~K. Sunori, P.~K. Juneja, M.~Chaturvedi, J.~Mittal, Dead time compensation in
  sugar crystallization process, in: Proceeding of International Conference on
  Intelligent Communication, Control and Devices, Springer, 2017, pp. 375--381.

\bibitem{ogunnaike1994process}
B.~Ogunnaike, W.~Ray,
  \href{https://books.google.com.ar/books?id=5LpAa9Le7ocC}{Process Dynamics,
  Modeling, and Control}, Topics in Chemical Engineering - Oxford University
  Press, Oxford University Press, 1994.
\newline\urlprefix\url{https://books.google.com.ar/books?id=5LpAa9Le7ocC}

\bibitem{dujovne20146tisch}
D.~Dujovne, T.~Watteyne, X.~Vilajosana, P.~Thubert, 6tisch: deterministic
  ip-enabled industrial internet (of things), IEEE Communications Magazine
  52~(12) (2014) 36--41.

\bibitem{RFC8655}
N.~Finn, P.~Thubert, B.~Varga, J.~Farkas,
  \href{https://rfc-editor.org/rfc/rfc8655.txt}{{Deterministic Networking
  Architecture}}, RFC 8655 (Oct. 2019).
\newblock \href {https://doi.org/10.17487/RFC8655}
  {\path{doi:10.17487/RFC8655}}.
\newline\urlprefix\url{https://rfc-editor.org/rfc/rfc8655.txt}

\bibitem{rfc6550}
R.~Alexander, A.~Brandt, J.~Vasseur, J.~Hui, K.~Pister, P.~Thubert, P.~Levis,
  R.~Struik, R.~Kelsey, T.~Winter,
  \href{https://rfc-editor.org/rfc/rfc6550.txt}{{RPL: IPv6 Routing Protocol for
  Low-Power and Lossy Networks}}, RFC 6550 (Mar. 2012).
\newblock \href {https://doi.org/10.17487/RFC6550}
  {\path{doi:10.17487/RFC6550}}.
\newline\urlprefix\url{https://rfc-editor.org/rfc/rfc6550.txt}

\bibitem{rfc8180}
X.~Vilajosana, K.~Pister, T.~Watteyne,
  \href{https://rfc-editor.org/rfc/rfc8180.txt}{{Minimal IPv6 over the TSCH
  Mode of IEEE 802.15.4e (6TiSCH) Configuration}}, RFC 8180 (May 2017).
\newblock \href {https://doi.org/10.17487/RFC8180}
  {\path{doi:10.17487/RFC8180}}.
\newline\urlprefix\url{https://rfc-editor.org/rfc/rfc8180.txt}

\bibitem{urke2021survey}
A.~R. Urke, {\O}.~Kure, K.~{\O}vsthus, A survey of 802.15. 4 tsch schedulers
  for a standardized industrial internet of things, Sensors 22~(1) (2021) 15.

\bibitem{rfc8480}
Q.~Wang, X.~Vilajosana, T.~Watteyne,
  \href{https://www.rfc-editor.org/info/rfc8480}{{6TiSCH Operation Sublayer
  (6top) Protocol (6P)}}, RFC 8480 (Nov. 2018).
\newblock \href {https://doi.org/10.17487/RFC8480}
  {\path{doi:10.17487/RFC8480}}.
\newline\urlprefix\url{https://www.rfc-editor.org/info/rfc8480}

\bibitem{rfc9033}
T.~Chang, M.~Vučinić, X.~Vilajosana, S.~Duquennoy, D.~R. Dujovne,
  \href{https://rfc-editor.org/rfc/rfc9033.txt}{{6TiSCH Minimal Scheduling
  Function (MSF)}}, RFC 9033 (May 2021).
\newblock \href {https://doi.org/10.17487/RFC9033}
  {\path{doi:10.17487/RFC9033}}.
\newline\urlprefix\url{https://rfc-editor.org/rfc/rfc9033.txt}

\bibitem{ldsf}
V.~Kotsiou, G.~Z. Papadopoulos, P.~Chatzimisios, F.~Theoleyre, {LDSF}:
  Low-latency distributed scheduling function for industrial internet of
  things, IEEE internet of things journal (2020).

\bibitem{resf}
G.~Daneels, B.~Spinnewyn, S.~Latr{\'e}, J.~Famaey, Resf: Recurrent low-latency
  scheduling in {IEEE} 802.15.4e {TSCH} networks, Ad Hoc Networks 69 (2018)
  100--114.

\bibitem{llsf}
T.~Chang, T.~Watteyne, Q.~Wang, X.~Vilajosana, Llsf: Low latency scheduling
  function for {6TiSCH} networks, in: 2016 International Conference on
  Distributed Computing in Sensor Systems (DCOSS), IEEE, 2016, pp. 93--95.

\bibitem{draft-ietf-6tisch-6top-sf0}
D.~R. Dujovne, L.~A. Grieco, M.~R. Palattella, N.~Accettura,
  \href{https://datatracker.ietf.org/doc/html/draft-ietf-6tisch-6top-sf0-05}{{6TiSCH
  6top Scheduling Function Zero (SF0)}}, Internet-Draft
  draft-ietf-6tisch-6top-sf0-05, Internet Engineering Task Force, work in
  Progress (Jul. 2017).
\newline\urlprefix\url{https://datatracker.ietf.org/doc/html/draft-ietf-6tisch-6top-sf0-05}

\bibitem{draft-ietf-raw-architecture}
P.~Thubert, G.~Z. Papadopoulos,
  \href{https://datatracker.ietf.org/doc/html/draft-ietf-raw-architecture-04}{{Reliable
  and Available Wireless Architecture}}, work in Progress (Mar. 2022).
\newline\urlprefix\url{https://datatracker.ietf.org/doc/html/draft-ietf-raw-architecture-04}

\bibitem{rfc9034}
L.~Thomas, S.~Anamalamudi, S.~Anand, M.~Hegde, C.~E. Perkins,
  \href{https://rfc-editor.org/rfc/rfc9034.txt}{{Packet Delivery Deadline Time
  in the Routing Header for IPv6 over Low-Power Wireless Personal Area Networks
  (6LoWPANs)}}, RFC 9034 (Jun. 2021).
\newblock \href {https://doi.org/10.17487/RFC9034}
  {\path{doi:10.17487/RFC9034}}.
\newline\urlprefix\url{https://rfc-editor.org/rfc/rfc9034.txt}

\bibitem{thubert2016ipv6}
P.~Thubert, R.~Cragie, Ipv6 over low-power wireless personal area network
  (6lowpan) paging dispatch, IETF, RFC 8025 (2016).

\bibitem{municio_simulating_2019}
E.~Municio, G.~Daneels, M.~Vučinić, S.~Latré, J.~Famaey, Y.~Tanaka, K.~Brun,
  K.~Muraoka, X.~Vilajosana, T.~Watteyne,
  \href{https://onlinelibrary.wiley.com/doi/abs/10.1002/ett.3494}{Simulating
  6tisch networks}, Transactions on Emerging Telecommunications Technologies
  30~(3) (2019) e3494.
\newblock \href {https://doi.org/10.1002/ett.3494}
  {\path{doi:10.1002/ett.3494}}.
\newline\urlprefix\url{https://onlinelibrary.wiley.com/doi/abs/10.1002/ett.3494}

\bibitem{leonardi2018multi}
L.~Leonardi, G.~Patti, L.~L. Bello, Multi-hop real-time communications over
  bluetooth low energy industrial wireless mesh networks, IEEE Access 6 (2018)
  26505--26519.

\bibitem{leonardi2023mrt}
L.~Leonardi, L.~L. Bello, G.~Patti, Mrt-lora: A multi-hop real-time
  communication protocol for industrial iot applications over lora networks,
  Computer Communications 199 (2023) 72--86.

\bibitem{chang20206tisch}
T.~Chang, M.~Vu{\v{c}}ini{\'c}, X.~V. Guill{\'e}n, D.~Dujovne, T.~Watteyne,
  6tisch minimal scheduling function: Performance evaluation, Internet
  Technology Letters 3~(4) (2020) e170.

\bibitem{hauweele2020pushing}
D.~Hauweele, R.-A. Koutsiamanis, B.~Quoitin, G.~Z. Papadopoulos, Pushing 6tisch
  minimal scheduling function (msf) to the limits, in: 2020 IEEE Symposium on
  Computers and Communications (ISCC), IEEE, 2020, pp. 1--7.

\bibitem{righetti2018analysis}
F.~Righetti, C.~Vallati, G.~Anastasi, S.~K. Das, Analysis and improvement of
  the on-the-fly bandwidth reservation algorithm for 6tisch, in: 2018 IEEE 19th
  International Symposium on" A World of Wireless, Mobile and Multimedia
  Networks"(WoWMoM), IEEE, 2018, pp. 1--9.

\bibitem{lim1989generalized}
K.~Lim, K.~Ling, Generalized predictive control of a heat exchanger, IEEE
  Control Systems Magazine 9~(6) (1989) 9--12.

\end{thebibliography}





\end{document}